\documentclass{easychair}

\usepackage{doc}
\usepackage{tikz}
\usepackage{float}
\usetikzlibrary{arrows.meta,backgrounds,calc,fit,positioning}
\definecolor{rowblue}{HTML}{DCEAF7}
\definecolor{rowblueborder}{HTML}{3973A5}
\definecolor{colgreen}{HTML}{E2F0E7}
\definecolor{colgreenborder}{HTML}{3E7C59}
\definecolor{accentgold}{HTML}{C17D11}
\definecolor{repairred}{HTML}{A23B3B}
\definecolor{ink}{HTML}{20242A}
\definecolor{muted}{HTML}{606A76}

\usepackage[disable]{todonotes}
\usepackage{siunitx}
\newcommand\occ{\ensuremath{\mathit{occ}}}
\newcommand{\solvername}[1]{\texttt{#1}}
\newcommand{\msolvername}[1]{\ensuremath{\mathtt{#1}}}
\newcommand{\primo}{\solvername{primo}}
\newcommand{\opensmt}{\solvername{opensmt}}
\newcommand{\primobest}{\msolvername{primo}\ensuremath{^{\star}}}
\newcommand{\primocomp}{\msolvername{primo}\ensuremath{^{+}}}

\title{Vibe-Coded and Tuned:\\
A State-of-the-Art SMT Solver for QF-LRA}
\author{Mikol{\'{a}}\v{s} Janota \and Jan Jakub\r{u}v}
\institute{
  Czech Technical University, Czechia\\
  \email{mikolas.janota@cvut.cz, jan.jakubuv@cvut.cz}
 }
\authorrunning{Janota and Jakub\r{u}v}
\titlerunning{QF-LRA: Vibe-Coded and Tuned}

\begin{document}
\maketitle

\begin{abstract}
This paper presents the SMT solver \primo{}, which is fully vibe-coded and then
parameter-tuned, achieving state-of-the-art results on linear real arithmetic
(QF-LRA). The performance of \primo{} is achieved by a systematic literature survey,
repeated profiling, and parameter tuning.
The resulting solver outperforms the winner of the QF-LRA track of
SMT-COMP~2026. This confirms that vibe-coding of automated reasoning tools will
enable us to make great strides in the future.
\end{abstract}

\section{Introduction}\label{sec:introduction} %
Agents powered by large language models (LLMs) have been undergoing rapid
development. It quickly became clear that such agents are capable of developing
formal methods tools~\cite{martins2026llmsbuildmaxsatsolver,arie2026} including SMT solvers~\cite{llm2smt}.
However, progress now in 2026 is so fast that this is no longer news. The
question that we pose here is whether they can also develop
\emph{state-of-the-art} SMT solvers.

In this paper we respond to this question positively for the theory of
quantifier-free linear real arithmetic (QF-LRA); we also give an indication
that similar results can be achieved for more complex theories, namely the
combination of QF-LRA and EUF, i.e.\ QF-UFLRA.

\begin{figure}[t]
  \centering
  \definecolor{workflowblue}{HTML}{255C8A}
  \definecolor{workflowteal}{HTML}{237A76}
  \definecolor{workflowviolet}{HTML}{66558C}
  \definecolor{workfloworange}{HTML}{A85D24}
  \definecolor{workflowink}{HTML}{263238}
  \definecolor{workflowpaper}{HTML}{F5F7F8}
  \begin{tikzpicture}[
      font=\sffamily,
      >={Stealth[length=2.2mm,width=1.5mm]},
      process/.style={
        rounded corners=2pt,
        minimum width=2.90cm,
        minimum height=1.72cm,
        text width=2.58cm,
        align=center,
        inner xsep=5pt,
        inner ysep=5pt,
        text=white,
        font=\sffamily\small
      },
      flow/.style={-{Stealth[length=2.2mm,width=1.5mm]},
        draw=workflowink, line width=0.75pt},
      record/.style={-{Stealth[length=1.8mm,width=1.3mm]},
        draw=workflowink!58, line width=0.6pt, densely dashed},
      edge label/.style={font=\sffamily\scriptsize, text=workflowink,
        fill=white, inner sep=1.5pt},
    ]

    \node[process, fill=workflowblue] (survey) at (0,0) {
      \textbf{LLM literature\\[-2pt]survey}\\[-1pt]
      \scriptsize Find algorithms, heuristics,\\[-1pt]
      \scriptsize and prior results
    };
    \node[process, fill=workflowteal] (code) at (3.40,0) {
      \textbf{LLM-assisted\\[-2pt]coding}\\[-1pt]
      \scriptsize Implement the solver from\\[-1pt]
      \scriptsize the surveyed literature
    };
    \node[process, fill=workflowviolet] (evaluate) at (6.80,0) {
      \textbf{Empirical\\[-2pt]evaluation}\\[-1pt]
      \scriptsize Profiling \enspace$\bullet$\enspace Fuzzing\\[-1pt]
      \scriptsize Benchmarking
    };
    \node[process, fill=workfloworange] (improve) at (10.20,0) {
      \textbf{Performance\\[-2pt]improvement}\\[-1pt]
      \scriptsize Tune algorithms, code,\\[-1pt]
      \scriptsize and parameters\\[-1pt]
      \scriptsize using evidence
    };

    \draw[flow] (survey.east) -- (code.west);
    \draw[flow] (code.east) -- (evaluate.west);
    \draw[flow] (evaluate.east) -- (improve.west);

    \draw[flow, rounded corners=7pt]
      (improve.north) -- ++(0,0.72) coordinate (return-east)
      -- node[edge label] {new questions and techniques} (return-east -| survey.north)
      -- (survey.north);

    \node[
      rounded corners=2pt,
      draw=workflowink!48,
      fill=workflowpaper,
      line width=0.65pt,
      minimum width=13.10cm,
      minimum height=1.30cm,
      text width=12.70cm,
      align=center,
      text=workflowink,
      font=\sffamily\small
    ] (docs) at (5.10,-2.12) {
      \textbf{Automatic documentation in Markdown (\texttt{*.md})}\\[-1pt]
      \scriptsize Literature notes \enspace$\bullet$\enspace Design rationale
      \enspace$\bullet$\enspace Code changes\\[-1pt]
      \scriptsize Experiment logs \enspace$\bullet$\enspace Benchmark results
    };

    \coordinate (doc-survey) at ($(docs.north west)!0.12!(docs.north east)$);
    \coordinate (doc-code) at ($(docs.north west)!0.37!(docs.north east)$);
    \coordinate (doc-evaluate) at ($(docs.north west)!0.63!(docs.north east)$);
    \coordinate (doc-improve) at ($(docs.north west)!0.88!(docs.north east)$);
    \draw[record] (survey.south) -- (doc-survey);
    \draw[record] (code.south) -- (doc-code);
    \draw[record] (evaluate.south) -- node[edge label, pos=0.56] {record} (doc-evaluate);
    \draw[record] (improve.south) -- (doc-improve);

    \begin{scope}[on background layer]
      \node[
        fit=(survey)(improve)(return-east)(docs),
        rounded corners=4pt,
        draw=workflowblue!62,
        fill=workflowblue!3,
        line width=0.85pt,
        densely dashed,
        inner xsep=8pt,
        inner ysep=8pt
      ] (supervision) {};
    \end{scope}
    \node[
      anchor=west,
      fill=white,
      text=workflowblue,
      inner xsep=4pt,
      font=\sffamily\footnotesize\bfseries
    ] at ([xshift=5pt]supervision.north west) {
      HUMAN SUPERVISION THROUGHOUT \normalfont--- direction, review, and validation
    };
  \end{tikzpicture}
  \vspace{-2pt}
  \caption{Our human-supervised, LLM-driven solver-development loop.  Each
    iteration connects the literature to implementation and empirical evidence;
    all stages continuously update a shared Markdown record.}
  \label{fig:solver-development-cycle}
\end{figure}
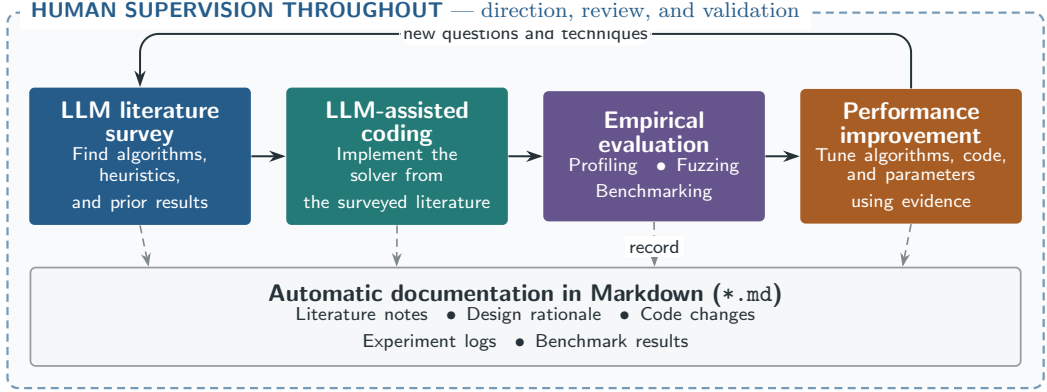

\autoref{fig:solver-development-cycle} summarizes the iterative,
human-supervised development process used in this work. We emphasize
sourcing from appropriate literature. Indeed, this seems to be the most
attractive skill we get from LLMs: the ability to understand an idea and
bring it to life by writing code. During development, we instructed the agent
to follow a \emph{clean-room policy} by not looking at the source code of other SMT
solvers; it may, however, look at the performance and overall statistics
provided by a reference solver. {\it The process needs to be iterated and guided
by a human}: the agent happily ignores parts of the literature it has found or
been given, or fails to follow it carefully. Intuition is often needed to pick
the next target; it therefore pays off to let the agent work on a well-defined
task and then seek human input.

Standard practices are a must: the agent constantly tests the solver by
fuzzing~\cite{fuzzing,Brummayer2009,winterer-zhang-su-oopsla2020}; it uses
delta-debugging tools to speed up bug analysis~\cite{ddsmt,Zeller2002}; it
routinely profiles with tools such as \texttt{gprof} to identify bottlenecks;
it uses the SMT-LIB benchmarks~\cite{smtlib} to assess progress; and it
documents its steps in plain English, in the form of Markdown files.

Our contributions are as follows.%
\footnote{We worked with the Claude Code agent (Opus~5 and Sonnet~5
models) and OpenAI Codex (GPT-5.4-mini, GPT-5.5, and GPT-5.6-sol). We write \emph{the agent}
throughout the paper, without distinguishing between them.}
\begin{itemize}
\item We present \primo{}, an SMT solver for QF-LRA (with preliminary
  support for QF-UFLRA) whose implementation was entirely vibe-coded, i.e.\
  written by an LLM-powered agent under human supervision and a clean-room
  policy that forbids consulting the source code of existing solvers.
\item We describe a human-supervised development methodology for building
  solvers this way, combining literature-driven target selection with
  standard solver-engineering practice --- fuzzing, delta-debugging,
  profiling, and SMT-LIB-based regression testing --- steered by iterative
  human feedback.
\item We present RamParILS, an LLM-designed solver configuration system and a
   parallelized rewrite of the iterated local search system
   ParamILS~\cite{paramils}. We carry out a systematic parameter-tuning
   study of \primo{}, producing a best-performing configuration (\primobest{})
   and one selected for complementarity with it (\primocomp{}).  We show that
   tuning \emph{alone} lifts \primo{} from below to above the state-of-the-art
   solvers \opensmt{} and \solvername{yices}.
\item We give an experimental evaluation on the full SMT-LIB QF-LRA
  benchmark set and on the SMT-COMP~2026 QF-LRA selection. This shows that
  \primobest{} solves more instances than default \opensmt{} at a $30$\,s
  timeout, with most of the gain concentrated in a single benchmark family,
  and that at the SMT-COMP \num{1200}\,s timeout it still improves
  PAR2 by $28\%$ over default \primo{} and $13\%$ over default \opensmt{}.
  This pattern is also observed on the instances held out from tuning.
\end{itemize}

\section{Preliminaries}\label{sec:preliminaries} %
We assume that the reader is familiar with classical first-order logic with
equality~\cite{smullyan}.
For the remainder of the paper we assume the standard DPLL(T) approach to
solving SMT~\cite{nieuwenhuis-jacm06}. We consider a quantifier-free linear
arithmetic (QF-LRA) formula a quantifier-free first-order formula whose atoms
are either propositional variables or equalities, disequalities, or inequalities
of the form \[a_1 x_1 + \dots + a_n x_n \bowtie b, \] where $a_1,\dots,a_n$ and
$b$ are rational numbers, $x_1\dots x_n$ are real variables, and $\bowtie$ is
one of the operators $\approx$, $\leq$, $<$ with standard syntactic sugar
operators $>$, $\geq$, $t-s$~\cite{kroening16}.

A \emph{theory solver} is responsible for accepting a set of literals in the given
theory and deciding whether it is satisfiable or not. If the set is unsatisfiable,
it is expected to provide an explanation which is a subset of the literals that
still is unsatisfiable. Following the DPLL(T) paradigm, the theory solver is
also responsible for propagation of values and may influence the decisions of
the SAT solver.

\subsection{The Solver}\label{sec:lra_in_primo} %
The solver \primo{} is written in C++ and as SAT solver it uses
CaDiCaL~\cite{cadical3}, which enables callbacks via the IPASIR-UP
interface~\cite{ipasirup}. For solving LRA, it is sufficient to consider only
rational numbers, but those need to be represented with arbitrary precision in
order to guarantee soundness; \primo{} uses the \texttt{GMP} library for this
purpose~\cite{gmp}. ANTLR~\cite{antlr} is used to parse the input. These
libraries were specified by human prompts to the agent from the beginning.

The development was started with Claude. The models used were Sonnet 5 and
Opus 5 (the latter released during the project's development). The QF-UFLRA
strand was mainly developed by Codex with the model gpt-5.6-sol.

Formulas are stored using the hash-consing mechanism (no formula is represented
twice in the memory). Interestingly, the agent decided not to implement
reference counting so no memory is released. This is likely not an issue in the
current setting because not much is discarded during solving, mainly during the
preprocessing phase. However, this might need to be reconsidered if we were to
support quantifiers. The current version has around 35k~LOC and another 14k~LOC
of tests.

Before solving, preprocessing is applied, which involves standard simplification
techniques as well as Gaussian elimination for top-level asserted equalities.
Since preprocessing is typically not idempotent, it is run repeatedly bounded by
a tunable parameter.

Only single query problems are accepted and there is support for
theories of uninterpreted functions (QF-UF), linear real arithmetic
(QF-LRA), and their combination (QF-UFLRA). Here we report mainly on QF-LRA as
QF-UFLRA is currently immature. The implementation of QF-UF is heavily
influenced by a previous incarnation of the solver LLM2SMT~\cite{llm2smt};
additionally, we observed that adding symmetries speeds up the QF-UF SMT-LIB
benchmarks~\cite{symmetries}. The combination QF-UFLRA is solved by
Nelson-Oppen closure (for convex theories) based on a model-guided approach
proposed by de Moura and Bj{\o}rner~\cite{MouraB08};
its performance is on par with \solvername{cvc5} but much slower than
\solvername{yices} (see \autoref{fig:uflra}). The solver does not yet
incorporate the improvements proposed by Bj{\o}rner and
Nachmanson~\cite{BjornerN24}.

The solver has two separate algorithms, one for QF-LRA and one for QF-UFLRA.
It is plausible that in the future QF-LRA formulas could just be solved by
QF-UFLRA as long as it is ensured that no additional overhead is incurred.
Nevertheless, at present we support the flag \texttt{--mixed-dispatch},
which routes QF-LRA to the QF-UFLRA algorithm whenever there are Boolean
constants in the problem; this has been useful for the diversity of
the solver in portfolio mode (see \autoref{sec:experiments}).

\subsection{LRA Theory Solver in primo}\label{sec:th_lra_in_primo}
The architecture of the theory solver is anchored in a simplex variant
introduced by Dutertre and de Moura~\cite{DutertreM06}. Unlike the traditional
simplex, this is geared towards the decision problem and supports lower and upper bounds for
variables. In its essence, it is a dual-simplex, since it gradually repairs
basic variables that are out of bounds.

A strong point of this approach is that backtracking is
cheap: no further pivoting is needed, values just need to be restored from
the stack. This is why this architecture was chosen. The theory solver only accepts
$\le$ and $<$ predicates. Equality $s \approx t$ is desugared into $s\leq t\land
t\leq s$ at the very beginning of solving. However, equalities may be propagated
during preprocessing, before the DPLL(T) loop begins. Following the original
paper, strict inequalities are treated by
$\delta$-rationals~\cite[Sec.~5]{DutertreM06}.

The original design of the simplex algorithm by Dutertre and de Moura chooses
the pivot by \emph{Bland's rule}~\cite{Bland1977} in order to
guarantee termination. Primo enables other pivoting rules but always falls back
to Bland's rule after a certain number of iterations. The number of iterations
after which we always use Bland's rule is $kn+c$, where $k$ defaults to $4$ and $c$ to
$64$. Both Griggio~\cite{griggio2009} and King~\cite{king2014} report on such
fallback techniques in their respective PhD theses.

\emph{Dantzig-style}: choose most-violated basic variable, then eligible
entering variable with largest absolute row coefficient~\cite[Sec.~2.2.12]{king2014}.
\emph{SOI-Simplex} globally minimizes the sum of
infeasibilities~\cite{King-fmcad2013}.

\emph{Sparse} pivot rule aims to limit \emph{fill-in} (the creation of new
nonzero tableau coefficients during pivoting). It first chooses a violated basic
variable whose defining row has the fewest nonzero coefficients, breaking ties
in favor of the largest bound violation. Among the nonbasic variables that can
repair this violation, it then chooses one occurring in the fewest tableau rows,
preferring a coefficient of $+1$ or $-1$ on ties. This favors pivots that
substitute a short row into as few other rows as possible. Minimum-column
entering selection and the finite switch to Bland's rule are standard heuristics
surveyed by King~\cite{king2014}. Monniaux motivates sparsity-oriented ordering
using both short defining rows and variables with few
occurrences~\cite{Monniaux2009}. Our shortest-row-first leaving order and its
sequential combination with minimum-column entering selection is a fusion of the
two. The Sparse pivot rule is the current default.

During the development, in collaboration with the authors, the agent proposed
 and implemented the following heuristics and improvements.

\paragraph{Sharing slacks.}
Given two atoms $0\leq x$, $x\leq 100$, the original implementation of
the algorithm would introduce two slack variables $s_1$ and $s_2$, both representing
$x$ and one bounded from below and one from above. Under the direct bounds
improvement, no new slacks are created and $x$ is directly bounded from below
as well as above. Slacks are also shared for bounds on compound terms, e.g.,
$x+y\leq 6$ and $x+y\leq 7$ introduce a single slack with the possible bounds
$s\leq 6$ and $s\leq 7$. Note that slacks are allocated before the DPLL(T) loop
starts, and therefore we do not know which of the bounds will be selected for
the particular theory solver call.

\paragraph{Small-rationals~\cite{Monniaux2009}.} A small wrapper has also been created around
\texttt{GMP} so that ``large'' numbers do not need to be allocated if the numbers
can be stored in processor-level integers; by default, 128 bits are used to
perform calculations on ``small'' numbers. It was discovered by profiling that
in many cases the numbers needed are small but \texttt{GMP} requires memory
allocation nonetheless. Upon further inspection, we found that Monniaux reports
on this approach~\cite{Monniaux2009}.

\paragraph{Model-based phase selection.}
Throughout its lifetime, the algorithm maintains an assignment $\beta$ giving a
value to each variable $x$ in the tableau. If there is an atom $x<u$ and
$\beta(x)<u$, we prefer this atom to be asserted positively, because this
variable does not need to be repaired---repair means pivoting on that variable.
It is analogous if $\beta(x)\geq u$. This heuristic selects the phase of the
Boolean variable representing the bound without affecting the order in which
variables are chosen (VSIDS is used by default).

\paragraph{Monotone variable elimination.}
If a variable is only bounded from below or only bounded from above in the
original formula, it can be eliminated; this is akin to a trivial step in
Fourier-Motzkin elimination~\cite{schrijver1998}. The nontrivial part is
the reconstruction of the model values (``post-solving''~\cite{Andersen1995}), e.g.,
$z<x\land x+5<y\land x+10<y$ eliminates all literals, where $z$ and $y$ are
unbounded but $x$ is bounded. The post-solving process traverses the eliminated
atoms and first pins $x$ to $0$; it then bounds $z$ from above by $0$, bounds
$y$ from below by $5$, and bounds $y$ from below by $10$, resulting in a lower
bound of $10$ for $y$ and an upper bound of $0$ for $z$.

\paragraph{Row-based propagation~\cite{DutertreM06}.} In the tableau, each basic variable $x$ is
defined by a combination of the non-basic ones, e.g., $x=y+z$; the bounds of the
non-basic ones can be combined in order to derive new bounds on $x$. In our
example, the sum of lower bounds gives us a lower bound for $x$; analogously,
for the upper bounds. To curb the cost of the technique, two tunable
parameters are introduced: \texttt{--lra-row-propagation-max-row-size}, which
skips propagation for rows over this size, and
\texttt{--lra-row-propagation-max-fanout}, which limits how many other
variables the propagation is recursively applied to.

\paragraph{Tableau representation.}
Several improvements were gained by improving the representation of the
tableau. The rows are indexed by variables, where the rows of non-basic
variables are unused (the outer array is dense; each row's own contents are
stored sparsely, as described next).
Each row is then a sparse vector of coefficients.
Monniaux describes a vector-based representation~\cite{Monniaux2009},
while King's thesis~\cite{king2014} proposes linked lists.
This comparison was carried out by the authors independently of the
development process, after \primo{}'s implementation was already complete:
\solvername{cvc5} uses linked lists, while \opensmt{}, \solvername{z3}, and
\solvername{yices} use vectors, though with some differences from
\primo{}'s scheme.
A notable feature of our representation is that vectors are \emph{sorted} by
variable ID; a feature present in \opensmt{} but not the others.
The second notable feature is how we keep track of in which row a variable
occurs.
After some experiments, we decided to \emph{not} use hash-sets to store this
piece of information but it is again a vector of vectors $\occ$,
i.e.\ $\occ[v]$ is the vector of rows in which $v$ appears;
this vector is unsorted, so removal is done by swap-pop.
Updates to the occurrence vectors were initially expensive but we realized that update is only really
necessary when the variable gets canceled out (its coefficient sums to 0);
for the entering variable $v$, $\occ[v]$ is emptied unconditionally.
Further, the tableau contains pointers back to the occurrence list, i.e.\ the
rows of the tableau contain triples $\langle x,c,i \rangle$, where $x$ is the
variable, $c$ its coefficient in the row, and $i$ the index in the $\occ[x]$
vector. This maintains the invariant that if a row indexed by variable $y$
contains the entry $\langle x,c,i \rangle$, then $\occ[x][i]=y$.
\solvername{yices} and \solvername{z3} maintain similar backlinks.
See also \autoref{sec:tableau} for illustrative picture.

\section{Parameter Vibe-Tuning}\label{sec:_parameter_tuning} %

SMT solvers typically expose a large number of options that control their
behavior, either given on the command line or set from within the SMT-LIB
input.  A solver ships with a default configuration that performs reasonably
well across the board. Choosing option values that make it perform well on a
particular benchmark is the \emph{algorithm configuration} problem,
which is well studied in the
literature~\cite{paramils,smac,ZhangMJK25,AleksandrovaJK24,LuSJDM024}. In this
section we describe our tuning of \primo{}'s options with
RamParILS,\footnote{\url{https://deeper4ai.github.io/ramparils}} our
vibe-coded Rust rewrite of the algorithm configuration tool
ParamILS~\cite{paramils}.  RamParILS extends ParamILS chiefly with parallel
evaluation and a fast result cache.  The rest of this section describes
RamParILS, the search it performs, and the parameter space we built for
\primo{}. The best configurations it finds for quantifier-free LRA problems from SMTLIB are evaluated in
\autoref{sec:experiments}.

\paragraph{Iterated Local Search (ILS).}  

The core of the ParamILS and RamParILS algorithms is iterated local search 
interleaved with random search to escape local optima~\cite{paramils}.
The input to the algorithm is a configuration space and a target benchmark,
and the goal is to find a configuration that performs best on that benchmark
under a given evaluation metric.  A configuration space is a set of parameters,
each with a finite set of values, and a \emph{configuration} $\theta$ assigns
one value to every parameter.

\begin{figure}[t]
  \centering
  \definecolor{ilsblue}{HTML}{255C8A}
  \definecolor{ilsteal}{HTML}{237A76}
  \definecolor{ilsviolet}{HTML}{66558C}
  \definecolor{ilsorange}{HTML}{A85D24}
  \definecolor{ilsink}{HTML}{263238}
  \definecolor{ilspaper}{HTML}{F5F7F8}
  {\hyphenpenalty=10000 \exhyphenpenalty=10000
  \begin{tikzpicture}[
    yscale=0.9,
      font=\sffamily,
      >={Stealth[length=2.2mm,width=1.5mm]},
      op/.style={
        rounded corners=2pt, align=center, text=white,
        inner xsep=5pt, inner ysep=4pt, font=\sffamily\footnotesize,
        text width=2.80cm, minimum height=1.55cm
      },
      st/.style={
        rounded corners=2pt, align=center, draw=ilsink!45, fill=ilspaper,
        text=ilsink, line width=0.6pt, font=\sffamily\footnotesize,
        inner xsep=5pt, inner ysep=4pt,
        text width=2.45cm, minimum height=1.55cm
      },
      flow/.style={-{Stealth[length=2.2mm,width=1.5mm]}, draw=ilsink,
        line width=0.75pt},
      wire/.style={draw=ilsink, line width=0.75pt},
      leg/.style={align=left, text=ilsink, font=\sffamily\scriptsize,
        text width=3.85cm, inner sep=0pt},
      lbl/.style={font=\sffamily\scriptsize, text=ilsink, fill=white,
        inner sep=1.5pt, align=center},
      band/.style={rounded corners=4pt, line width=0.75pt, densely dashed,
        inner xsep=7pt, inner ysep=9pt},
      bandlbl/.style={font=\sffamily\scriptsize\bfseries, text=ilsink,
        fill=white, inner xsep=3pt, anchor=west},
    ]

    \node[st] (t0) at (0,0) {
      $\theta$\\[1pt]
      \scriptsize\sffamily input config\\[-1pt]
      \scriptsize\sffamily (or random)
    };
    \node[op, fill=ilsteal] (ev0) at (3.55,0) {
      \textbf{Evaluate}\\[1pt]
      \scriptsize\sffamily the input config $\theta$\\[-1pt]
      \scriptsize\sffamily on all instances
    };

    \node[op, fill=ilsblue] (bls0) at (7.75,0) {
      \textbf{Basic local search}\\[1pt]
      \scriptsize\sffamily explores single-parameter
      neighbors of $\theta$\\[2pt]
      \scriptsize\sffamily $\theta_{\mathit{base}} \leftarrow$ local optimum
    };
    \node[st] (tils) at (11.15,0) {
      $\theta_{\mathit{base}}$\\[1pt]
      \scriptsize\sffamily initial local optimum\\[-1pt]
      \scriptsize\sffamily (perturbation base)\\[2pt]
      \scriptsize\sffamily $\theta_{\mathit{inc}} \leftarrow \theta_{\mathit{base}}$
    };

    \draw[flow] (t0.east) -- (ev0.west);
    \draw[flow] (ev0.east) -- (bls0.west);
    \draw[flow] (bls0.east) -- (tils.west);

    \node[op, fill=ilsviolet, minimum height=1.70cm] (pert) at (0.35,-3.75) {
      \textbf{Perturbation}\\[1pt]
      \scriptsize\sffamily $s$ single-parameter\\[-1pt]
      \scriptsize\sffamily random flips from $\theta_{\mathit{base}}$\\[2pt]
      \scriptsize\sffamily $\theta \leftarrow$ perturbed config
    };
    \node[op, fill=ilsteal, minimum height=1.70cm] (evp) at (3.90,-3.75) {
      \textbf{Evaluate}\\[1pt]
      \scriptsize\sffamily the perturbed config $\theta$\\[-1pt]
      \scriptsize\sffamily on all instances
    };
    \node[op, fill=ilsblue, minimum height=1.70cm] (blsp) at (7.45,-3.75) {
      \textbf{Basic local search}\\[1pt]
      \scriptsize\sffamily explores single-parameter
      neighbors of $\theta$\\[2pt]
      \scriptsize\sffamily $\theta \leftarrow$ local optimum\\[-1pt]
      \scriptsize\sffamily $\theta_{\mathit{inc}} \leftarrow \theta$ if better
    };
    \node[op, fill=ilsorange, minimum height=1.70cm] (accp) at (11.00,-3.75) {
      \textbf{Acceptance criterion}\\[1pt]
      \scriptsize\sffamily compares $\theta$ with $\theta_{\mathit{base}}$\\[2pt]
      \scriptsize\sffamily $\theta_{\mathit{base}} \leftarrow \theta$ if better
    };

    \draw[flow] (pert.east) -- (evp.west);
    \draw[flow] (evp.east) -- (blsp.west);
    \draw[flow] (blsp.east) -- (accp.west);

    \coordinate (corr) at (0.35,-2.05);
    \draw[wire] (tils.south) -- (11.15,-2.30);
    \draw[wire] ([xshift=1.45cm]accp.north) -- ++(0,0.675);
    \draw[flow, rounded corners=6pt] (12.45,-2.30)
      -- node[lbl] {$\theta_{\mathit{base}}$} (0.35,-2.30) -- (pert.north);
    \fill[ilsink] (11.15,-2.30) circle (1.5pt);

    \node[leg] (lg1) at (0.70,-5.55) {
      $\theta = {}$current config};
    \node[leg] (lg2) at (5.40,-5.55) {
      $\theta_{\mathit{base}} = {}$perturbation start config};
    \node[leg] (lg3) at (10.10,-5.55) {
      $\theta_{\mathit{inc}} = {}$incumbent: the best config};
    \begin{scope}[on background layer]
      \node[fit=(lg1)(lg2)(lg3), rounded corners=2pt, draw=ilsink!45,
        fill=ilspaper, line width=0.65pt, inner xsep=6pt, inner ysep=6pt] {};
    \end{scope}

    \begin{scope}[on background layer]
      \node[band, fit=(t0)(ev0), draw=ilsink!40, fill=ilsink!3] (binit) {};
      \node[band, fit=(bls0)(tils), draw=ilsink!40, fill=ilsink!3] (bfirst) {};
      \node[band, fit=(pert)(evp)(blsp)(accp)(corr),
        draw=ilsviolet!55, fill=ilsviolet!4] (bloop) {};
    \end{scope}
    \node[bandlbl] at ([xshift=6pt]binit.north west)
      {Basic ILS: initialization};
    \node[bandlbl] at ([xshift=6pt]bfirst.north west)
      {Basic ILS: first local search};
    \node[bandlbl] at ([xshift=6pt]bloop.north west)
      {Basic ILS: main loop \normalfont--- repeat until the deadline,
       then return $\theta_{\mathit{inc}}$};

  \end{tikzpicture}}
  \caption{Basic Iterative Local Search (ILS) implemented in
    ParamILS~\cite{paramils} and RamParILS.}
  \label{fig:basic-ils}
\end{figure}
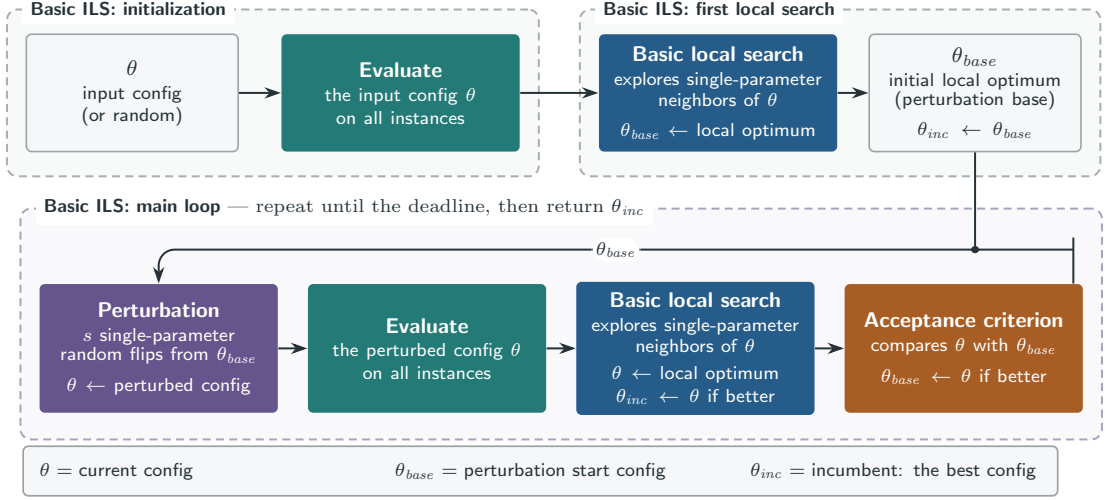
 
\autoref{fig:basic-ils} illustrates the \emph{basic} variant of the ILS
algorithm.  Starting from the input configuration $\theta$, the algorithm
repeatedly explores the neighboring configurations, those that differ in
exactly one parameter, and moves to the first one that scores better.  This
local search is interleaved with a \emph{perturbation} phase, in which $s$
parameters are set at random in an attempt to escape a local optimum, where
$s$ is called the \emph{perturbation strength}.  The
configuration a perturbation starts from is $\theta_{\mathit{base}}$, and it
is replaced whenever the acceptance criterion prefers the local optimum just
found.  The best configuration seen, the \emph{incumbent}
$\theta_{\mathit{inc}}$, is returned once the tuning time budget is exhausted.
A configuration is scored by the mean runtime of the solver over the tuning
instances.

Basic ILS always evaluates a configuration on all available instances.  The
alternative, \emph{focused ILS}, scores configurations on a growing prefix of
the instances, so that weak candidates are rejected after few solver runs.  We
do not need it: our tuning set is small enough to score every configuration on
all of it, which has the further advantage that all scores are directly
comparable.  Both variants are implemented in RamParILS.

\paragraph{RamParILS.}  ParamILS is an iterated local search over a discrete
parameter space, written in Ruby.  RamParILS is a reimplementation in Rust,
intended to speed up the tuning process through parallel configuration
evaluation and persistent result caching.  Its main extensions over ParamILS
are the following.
\begin{enumerate}
  \item \emph{Parallel evaluation.}  ParamILS evaluates one configuration at a
    time.  RamParILS submits a whole neighborhood to a pool of workers and
    takes the first neighbor that finishes with a better score, so that with
    $60$ workers an evaluation that would take an hour sequentially finishes
    in minutes.
  \item \emph{Persistent cache.}  Every solver run is stored in an SQLite
     database keyed by the configuration and the instance.  The database
     outlives the run, so a configuration that the search revisits costs no
     solver calls.  This matters because several RamParILS runs are typically
     launched during a tuning session.  Caching the instances that time out is
     especially valuable, since those are the most expensive to re-run.
  \item \label{itm:escape}\emph{Escaping a frozen search.}  ParamILS accepts a new local optimum
     only when it is at least as good as the current one, so a search that
     stops improving perturbs the same configuration for the rest of its
     budget.  ParamILS addresses this by random restarts under a
     fixed probability.  RamParILS in addition accepts a slightly worse
     local optimum, and it restarts from the best configuration found after
     several rejections in a row.

\end{enumerate}

RamParILS was itself written by an LLM agent.  The initial request was a parallel and
caching rewrite of ParamILS in Rust.  The design of the new features was
discussed in detail to ensure the efficiency of the implementation, while the
remaining ParamILS features were to be reproduced as in the original Ruby
version.  Extension~\ref{itm:escape} was not part of that plan and was added
along the way: two tuning runs turned out to spend most of their budget
perturbing the same configuration, and the escape mechanism was implemented in
response. The agent was able to carry out the project on its own, and it was
also used to analyze the experiment logs and to locate and fix bugs.

\paragraph{Parameter space.}  \primo{} exposes ca.~$90$
command-line options.  Most of them are opt-in heuristics that were
implemented, tested on a developer-scale sample, and assigned a reasonable
default.  From the options that plausibly affect the QF-LRA problems in our
benchmark we built a space of $11$ parameters and \num{8640} configurations,
in which \primo{}'s default configuration has $16$ neighbors.

The agent was able to analyze \primo{}'s source code and design the parameter
space on its own.  This helps close a usability gap in parameter tuning, where
setting up a space normally demands solver-specific knowledge.  It also shrank
the space considerably: from the logs of trial RamParILS runs and the results
in its cache, the agent detected dependencies between parameters, which
RamParILS can express by declaring a parameter active only for certain values
of another.  Such dependencies are invisible from the outside, showing up only
as flat regions of the space in which whole neighborhoods score identically. 
Finding them required an inspection of the source.

\paragraph{Tuning setting.}  
Scoring a configuration on the whole SMT-LIB QF-LRA
benchmark~\cite{smtlib} of \num{1753} instances can be too expensive, so we
tune on a representative subset of $473$ instances. Based on a preliminary evaluation of several
configurations, we remove trivial problems solved by all of them. We take
the $300$ instances with the highest variance across configurations, subject to
a floor of eight per benchmark family so that no family is dropped entirely.
Additionally, $151$ instances unsolved by all configurations are added in
proportion to each family's share.

We tune on this subset with a timeout of $10$ seconds per solver run, termed
\emph{cutoff}. The tuning objective is to minimize the mean runtime; we use
perturbation strength $s = 3$ and an overall tuning time budget of $12$ hours.
The runs are reported in \autoref{sec:experiments}.

\section{Experiments: Strategy Tuning and Analysis}\label{sec:experiments} %

We perform several tuning runs, evaluate the promising configurations on all
QF-LRA instances of SMT-LIB to select two \primo{} configurations, and compare
these in an experiment reproducing the SMT-COMP~2026 competition on QF-LRA
instances~\cite{smtcomp2026}.
We compare the \primo{} configurations with two state-of-the-art SMT solvers,
\opensmt{}~\cite{opensmt2} and \solvername{yices}~\cite{yices2}, which were the
two best solvers on QF-LRA at SMT-COMP~2026.  The \primo{} configurations
evaluated are (1) the default configuration, (2) \primobest{}, the best
configuration found by tuning, and (3) \primocomp{}, the configuration most
complementary to \primobest{}.%
\footnote{Experiments performed on two AMD EPYC 7513 32-core processors with
512\,GB of RAM, running $64$ solver instances concurrently.  We use
\opensmt{}~\texttt{v2.9.2-35-g15b42c6f} and \solvername{yices}~$2.7.0$
(revision \texttt{4f19700f}).}

\paragraph{Tuning runs.}

We made several RamParILS runs starting from different initial configurations.
The first run starts from \primo{}'s default configuration, while each
subsequent run restarts from the final configuration of the previous one.  From
the run logs, we extract all configurations reported as incumbents and evaluate
them on all QF-LRA instances from SMT-LIB. The overall best configuration,
denoted \primobest{}, and its most complementary configuration, denoted
\primocomp{}, are reported in \autoref{fig:strategies}.

What the tuner selected are switches that turn on techniques of
\autoref{sec:th_lra_in_primo}, each of which \primo{} leaves off by
default.  Of these, \primobest{} turns on the first two:
(1)~\texttt{lra-model-phase} enables model-based phase selection, and
(2)~\texttt{monotone-elimination} enables monotone variable elimination in
preprocessing.  The complementary configuration \primocomp{} is a proper
extension of \primobest{}, keeping both and adding four flags.  Flag~(3),
\texttt{mixed-dispatch wide}, counts a Bool-sorted constant as an EUF
predicate, routing pure QF-LRA formulas that contain propositional variables
through the combined EUF/LRA path.  The remaining three concern row-based
propagation: (4) enables it, while (5) and (6) widen its cost caps from the
default of $64$, to $256$ for the size of a row propagated over and to
\num{1024} for the fanout.

\begin{figure}[t!]
\centering
\small
\begin{tabular}{|l|l|}
\hline
\multicolumn{1}{|c|}{\primobest{}} & \multicolumn{1}{c|}{\primocomp{}} \\
\hline
\begin{tabular}[t]{@{}r@{~}l@{}}
1. & \texttt{--lra-model-phase}\\
2. & \texttt{--monotone-elimination}
\end{tabular}
&
\begin{tabular}[t]{@{}r@{~}l@{}}
1. & \texttt{--lra-model-phase}\\
2. & \texttt{--monotone-elimination}\\
3. & \texttt{--mixed-dispatch wide}\\
4. & \texttt{--lra-row-propagation}\\
5. & \texttt{--lra-row-propagation-max-row-size 256}\\
6. & \texttt{--lra-row-propagation-max-fanout 1024}
\end{tabular}
\\
\hline
\end{tabular}
\caption{The two best \primo{} configurations found by RamParILS tuning.}
\label{fig:strategies}
\end{figure}

\paragraph{Full QF-LRA evaluation.}
\autoref{tab:t30} reports the whole QF-LRA division of SMT-LIB, \num{1753}
instances.  Of these, SMT-LIB declares $946$ satisfiable and $683$
unsatisfiable, while the remaining $124$ declare an unknown status.  The
three status columns of Table~\ref{tab:t30} report solver performance on the
problems declared with that status.
Between them the five configurations settle $102$ of the $124$ open
instances, and $82$ instances are not solved at all. The overall winner is
\primobest{} solving \num{1599} instances in total and reaching the best PAR2
score (the sum of runtimes, where an unsolved instance counts as twice the
cutoff). Note that \opensmt{} maintains its superiority on unsatisfiable
instances as it keeps a $23$-instance lead even against \primobest{}.

\begin{table}[t]
\centering
\small
\setlength{\tabcolsep}{4pt}
\begin{tabular}{l|rr|rr|rr|rr|r}
\hline
configuration & \multicolumn{2}{c|}{total (\num{1753})}
  & \multicolumn{2}{c|}{sat ($946$)}
  & \multicolumn{2}{c|}{unsat ($683$)}
  & \multicolumn{2}{c|}{unknown ($124$)} & PAR2 \\
\hline
\primobest{}         & \textbf{\num{1599}} & \textbf{$91.2\%$}
  & 890 & $94.1\%$ & 614 & $89.9\%$ & \textbf{95} & \textbf{$76.6\%$}
  & \textbf{\num{12806}} \\
\opensmt{} (default) & \num{1578} & $90.0\%$
  & 853 & $90.2\%$ & \textbf{637} & \textbf{$93.3\%$} & 88 & $71.0\%$
  & \num{14193} \\
\solvername{yices} (default) & \num{1563} & $89.2\%$
  & 889 & $94.0\%$ & 580 & $84.9\%$ & 94 & $75.8\%$
  & \num{13886} \\
\primocomp{}         & \num{1562} & $89.1\%$
  & \textbf{892} & \textbf{$94.3\%$} & 583 & $85.4\%$ & 87 & $70.2\%$
  & \num{15458} \\
\primo{} (default)   & \num{1536} & $87.6\%$
  & 857 & $90.6\%$ & 585 & $85.7\%$ & 94 & $75.8\%$
  & \num{16597} \\
\hline
\emph{VBS (union)}   & \num{1671} & $95.3\%$
  & 922 & $97.5\%$ & 647 & $94.7\%$ & 102 & $82.3\%$
  & \num{7673} \\
\hline
\end{tabular}
\caption{Evaluation on full SMT-LIB QF-LRA (\num{1753} instances), $30$\,s
cutoff.}
\label{tab:t30}
\end{table}

The best tuned configuration \primobest{} improves over the default
\primo{} by $63$ instances, which is enough to outperform default \opensmt{}
by $21$ and \solvername{yices} by $36$.  The gain is not spread evenly over QF-LRA: $62$ of
the $77$ instances \primobest{} adds come
from the \texttt{LassoRanker} family, program-analysis queries encoded through
Farkas' lemma, whose large disjunctive Boolean structure is what
\texttt{monotone-elimination} simplifies.  %
Tuning for complementarity costs aggregate performance: \primocomp{} ends $37$
instances behind \primobest{}, yet solves $23$ that \primobest{} misses. 
The improvement of \primobest{} over the default \primo{} is not free: it
adds $77$ instances but loses $14$, and all the lost instances are satisfiable.
This is because the option \texttt{lra-model-phase} steers the polarity from
one candidate model, which can make the solver miss a model that lies
elsewhere.  On unsatisfiable instances there is no model to miss.

The most complementary pair is \primobest{} with default \opensmt{}: together
they solve \num{1639} instances, $40$ more than \primobest{} alone.  Pairing
\primobest{} with \solvername{yices} instead solves \num{1638}, $39$ more than
\primobest{} alone.  All three solvers combined reach \num{1660}, $61$ more
than \primobest{} alone, and all five configurations together \num{1671}, $72$
more.  The two reference solvers complement \primobest{} on opposite answers:
$28$ of the $40$ instances \opensmt{} adds are unsatisfiable, whereas $20$ of
the $39$ added by \solvername{yices} are satisfiable.

\paragraph{The SMT-COMP competition experiments.}

Since the annual SMT-COMP competition uses a much higher cutoff of \num{1200}
seconds, we repeat the comparison on the SMT-COMP~2026 QF-LRA
selection~\cite{smtcomp2026} at that cutoff.  The only difference from the
actual competition is the memory limit, which we set to $7$\,GB as opposed to
the $30$\,GB used there, so that all $64$ workers fit in memory in the worst
case.  This has negligible effect, since only a single solver run
terminated out of memory.
The results are presented in \autoref{tab:t1200}, which splits the $519$
instances into the $194$ that belong to the tuning set and the $325$ test
instances never used for tuning.

\begin{table}[t]
\centering
\small
\setlength{\tabcolsep}{4pt}
\begin{tabular}{l|rr|rr|rr|rrr}
\hline
 & \multicolumn{6}{c|}{solved} & \multicolumn{3}{c}{PAR2} \\
\cline{2-7}\cline{8-10}
configuration & \multicolumn{2}{c|}{all ($519$)}
  & \multicolumn{2}{c|}{train ($194$)}
  & \multicolumn{2}{c|}{test ($325$)} & all & train & test \\
\hline
\primobest{}         & \textbf{504} & \textbf{$97.1\%$}
  & \textbf{192} & \textbf{$99.0\%$} & \textbf{312} & \textbf{$96.0\%$}
  & \textbf{\num{47653}} & \textbf{\num{6509}} & \textbf{\num{41144}} \\
\primocomp{}         & 502 & $96.7\%$
  & \textbf{192} & \textbf{$99.0\%$} & 310 & $95.4\%$
  & \num{53167} & \num{6818} & \num{46350} \\
\opensmt{} (default) & 502 & $96.7\%$
  & \textbf{192} & \textbf{$99.0\%$} & 310 & $95.4\%$
  & \num{54584} & \num{8253} & \num{46331} \\
\primo{} (default)   & 499 & $96.1\%$
  & 191 & $98.5\%$ & 308 & $94.8\%$
  & \num{65821} & \num{11426} & \num{54395} \\
\solvername{yices} (default) & 495 & $95.4\%$
  & 187 & $96.4\%$ & 308 & $94.8\%$
  & \num{73077} & \num{19588} & \num{53489} \\
\hline
\emph{VBS (union)}   & 511 & $98.5\%$
  & 193 & $99.5\%$ & 318 & $97.8\%$
  & \num{25725} & \num{3400} & \num{22325} \\
\hline
\end{tabular}
\caption{Evaluation on the SMT-COMP~2026 QF-LRA selection ($519$ instances),
\num{1200}\,s cutoff.}
\label{tab:t1200}
\end{table}

Instance counts almost stop discriminating here: only $8$ of the $519$ are
unsolved by all five configurations, and the spread among the four
\primo{}/\opensmt{} rows is $5$, against $63$ at the $30$\,s cutoff.  The train
half is close to saturated, with default \primo{} already solving $98.5\%$ of
it, and this follows from how the subset was chosen rather than from the
tuning: its instances were selected because they discriminate between
configurations at a $10$\,s cutoff, and given \num{1200} seconds nearly all of
them come within reach of every solver.  What discrimination remains therefore
sits in the held-out half, where \primobest{} gains $4$ of its $5$ instances.  A
configuration that had overfitted the tuning subset would show the opposite
pattern.

The runtime does not seem to be saturated in this way since \primobest{}
improves PAR2 by $28\%$ against default \primo{} and by $13\%$ against default
\opensmt{}.
On the $483$ instances all five solve, it takes $46.6\%$ less total time than
default \primo{} (\num{7284}\,s against \num{13648}\,s), where a repeated
measurement of the same configuration suggests a noise level of about $1\%$.  The
instance-count advantage of tuning is largely a short-cutoff phenomenon while
the speedup is not.

\section{Conclusions and Future Work}\label{sec:conclusions_and_future_work} %
We describe a fully vibe-coded SMT solver for the theory of quantifier-free
linear arithmetic (QF-LRA).  We also study automated tuning of the solver's
parameters.\todo{Summarize the main quantitative findings here: the gains over
default \primo{} and \opensmt{} at $30$\,s, the PAR2 improvement at
\num{1200}\,s, the held-out result, and the concentration of the short-cutoff
gain in the \texttt{LassoRanker} family.} \opensmt{} was in fact the winner of
the QF-LRA (single query) track of SMT-COMP~2026, and \primobest{} outperforms
it both in the number of instances solved and in PAR2 score.

Our motivation is mainly to gauge how well
LLM-guided agents can integrate existing literature into code. One valid
concern is whether the agent simply copies code from existing solvers into the
new one. Our experience speaks against this: the agent behaves much like a
human, in the sense that it usually starts with a naive implementation of what
it understands and requires considerable human supervision and further effort
to make it efficient. Currently, the repository contains roughly one thousand
commits spanning one month (the agent performs commits on its own and is
instructed to do so frequently). We have also used
anti-plagiarism software, Dolos~\cite{dolos} and PMD~\cite{pmd}, run against
\solvername{z3}, \opensmt{}, \solvername{yices}, and \solvername{cvc5}.
This analysis did not yield any overlaps apart from boilerplate enums.
Nevertheless, this remains an inherent threat to validity,  because we cannot
control on which data the LLM was trained on.

The use of an agent helps us quickly integrate, evaluate, and eventually tune
various techniques, as shown in this paper. One natural issue arising from
integrating many different algorithms into the solver is correctness. We used
fuzzing during development, but it turned out to be insufficient in some cases
or insufficiently seeded. To address this, we have also included some
hand-written fuzzers for dedicated fragments, which have weeded out some
bugs---one such tricky fragment was functions taking Boolean arguments.
Increasing the test coverage further is subject to future work.

This work has enabled us to produce a solver that is state of the art in the
sense reported in \autoref{sec:experiments}: on the SMT-LIB QF-LRA benchmark
and the SMT-COMP~2026 QF-LRA selection, \primobest{} solves more instances and
reaches a better PAR2 score than the default configurations of \opensmt{} and
\solvername{yices} at both the $30$\,s and \num{1200}\,s cutoffs used there. We
also relate the implementation to the existing literature, which we hope is
useful to the community. Currently, we also have an implementation of Nelson--Oppen closure
combining real arithmetic and uninterpreted functions. Improving this
implementation and extending the solver to integers are therefore natural
continuations of this work.\todo{Also make stronger validation a primary item of
future work, following the correctness limitation above. Reconcile with the
QF-UFLRA results now shown in \autoref{fig:uflra}: the ``preliminary, not
evaluated'' framing used elsewhere no longer holds.}

\bibliographystyle{plainurl}

\clearpage
\appendix

\section{Performance Cactus Plots}

\begin{figure}[H]
\centering
\definecolor{cacA}{HTML}{237A76}
\definecolor{cacB}{HTML}{A85D24}
\definecolor{cacC}{HTML}{255C8A}
\definecolor{cacD}{HTML}{66558C}
\begin{tikzpicture}[font=\sffamily\scriptsize,
    ax/.style={draw=black!55, line width=0.6pt},
    gr/.style={draw=black!12, line width=0.4pt},
    gm/.style={draw=black!7, line width=0.3pt},
    cv/.style={line width=1.0pt, line join=round, line cap=round}]
  \node[font=\sffamily\small\bfseries] at (2.75,4.37) {full benchmark, $30$\,s};
  \draw[gr] (0.000,0) -- (0.000,3.95);
  \draw[gm] (0.476,0) -- (0.476,3.95);
  \draw[gm] (0.755,0) -- (0.755,3.95);
  \draw[gm] (0.952,0) -- (0.952,3.95);
  \draw[gm] (1.106,0) -- (1.106,3.95);
  \draw[gm] (1.231,0) -- (1.231,3.95);
  \draw[gm] (1.337,0) -- (1.337,3.95);
  \draw[gm] (1.428,0) -- (1.428,3.95);
  \draw[gm] (1.509,0) -- (1.509,3.95);
  \draw[gr] (1.582,0) -- (1.582,3.95);
  \draw[gm] (2.058,0) -- (2.058,3.95);
  \draw[gm] (2.336,0) -- (2.336,3.95);
  \draw[gm] (2.534,0) -- (2.534,3.95);
  \draw[gm] (2.687,0) -- (2.687,3.95);
  \draw[gm] (2.813,0) -- (2.813,3.95);
  \draw[gm] (2.919,0) -- (2.919,3.95);
  \draw[gm] (3.010,0) -- (3.010,3.95);
  \draw[gm] (3.091,0) -- (3.091,3.95);
  \draw[gr] (3.164,0) -- (3.164,3.95);
  \draw[gm] (3.640,0) -- (3.640,3.95);
  \draw[gm] (3.918,0) -- (3.918,3.95);
  \draw[gm] (4.116,0) -- (4.116,3.95);
  \draw[gm] (4.269,0) -- (4.269,3.95);
  \draw[gm] (4.394,0) -- (4.394,3.95);
  \draw[gm] (4.500,0) -- (4.500,3.95);
  \draw[gm] (4.592,0) -- (4.592,3.95);
  \draw[gm] (4.673,0) -- (4.673,3.95);
  \draw[gr] (4.745,0) -- (4.745,3.95);
  \draw[gm] (5.221,0) -- (5.221,3.95);
  \draw[gm] (5.500,0) -- (5.500,3.95);
  \draw[gr] (0.00,0.000) -- (5.50,0.000);
  \node[anchor=east] at (-0.08,0.000) {50};
  \draw[gr] (0.00,0.790) -- (5.50,0.790);
  \node[anchor=east] at (-0.08,0.790) {60};
  \draw[gr] (0.00,1.580) -- (5.50,1.580);
  \node[anchor=east] at (-0.08,1.580) {70};
  \draw[gr] (0.00,2.370) -- (5.50,2.370);
  \node[anchor=east] at (-0.08,2.370) {80};
  \draw[gr] (0.00,3.160) -- (5.50,3.160);
  \node[anchor=east] at (-0.08,3.160) {90};
  \draw[gr] (0.00,3.950) -- (5.50,3.950);
  \node[anchor=east] at (-0.08,3.950) {100};
  \draw[ax] (0.00,0) -- (0.00,-0.09);
  \node[anchor=north] at (0.00,-0.12) {0.01};
  \draw[ax] (1.58,0) -- (1.58,-0.09);
  \node[anchor=north] at (1.58,-0.12) {0.1};
  \draw[ax] (3.16,0) -- (3.16,-0.09);
  \node[anchor=north] at (3.16,-0.12) {1};
  \draw[ax] (4.75,0) -- (4.75,-0.09);
  \node[anchor=north] at (4.75,-0.12) {10};
  \draw[ax] (5.50,0) -- (5.50,-0.09);
  \node[anchor=north] at (5.50,-0.12) {30};
  \draw[ax] (0.00,0) -- (5.50,0);
  \draw[ax] (0.00,0) -- (0.00,3.95);
  \node[anchor=north] at (2.75,-0.52) {time (s)};
  \node[rotate=90, anchor=south] at (-0.88,1.98) {solved [\%]};
  \begin{scope}
  \clip (0.00,-0.02) rectangle (5.50,4.00);
  \draw[cv, cacA, densely dashed] (0.000,-0.818) -- (0.075,-0.759) -- (0.150,-0.728) -- (0.200,-0.696) -- (0.250,-0.651) -- (0.300,-0.633) -- (0.325,-0.611) -- (0.375,-0.597) -- (0.425,-0.557) -- (0.450,-0.525) -- (0.500,-0.493) -- (0.525,-0.480) -- (0.550,-0.462) -- (0.575,-0.439) -- (0.625,-0.421) -- (0.650,-0.403) -- (0.675,-0.363) -- (0.700,-0.354) -- (0.725,-0.340) -- (0.750,-0.327) -- (0.775,-0.322) -- (0.800,-0.277) -- (0.825,-0.264) -- (0.850,-0.250) -- (0.875,-0.241) -- (0.900,-0.228) -- (0.925,-0.205) -- (0.950,-0.196) -- (0.975,-0.164) -- (1.000,-0.151) -- (1.025,-0.128) -- (1.050,-0.119) -- (1.075,-0.115) -- (1.100,-0.106) -- (1.125,-0.092) -- (1.150,-0.083) -- (1.175,-0.052) -- (1.200,-0.043) -- (1.225,-0.029) -- (1.250,-0.016) -- (1.275,0.002) -- (1.300,0.011) -- (1.325,0.016) -- (1.350,0.034) -- (1.375,0.043) -- (1.400,0.088) -- (1.425,0.110) -- (1.450,0.128) -- (1.475,0.142) -- (1.500,0.160) -- (1.525,0.178) -- (1.550,0.187) -- (1.575,0.223) -- (1.600,0.228) -- (1.625,0.241) -- (1.650,0.250) -- (1.675,0.255) -- (1.700,0.277) -- (1.725,0.300) -- (1.750,0.318) -- (1.775,0.336) -- (1.800,0.358) -- (1.825,0.381) -- (1.850,0.394) -- (1.875,0.408) -- (1.925,0.412) -- (1.950,0.426) -- (1.975,0.444) -- (2.000,0.466) -- (2.025,0.489) -- (2.050,0.507) -- (2.075,0.521) -- (2.100,0.548) -- (2.125,0.588) -- (2.150,0.606) -- (2.175,0.629) -- (2.200,0.633) -- (2.225,0.651) -- (2.250,0.669) -- (2.275,0.683) -- (2.300,0.710) -- (2.325,0.750) -- (2.350,0.764) -- (2.375,0.782) -- (2.400,0.795) -- (2.425,0.813) -- (2.450,0.827) -- (2.475,0.836) -- (2.500,0.854) -- (2.525,0.868) -- (2.550,0.904) -- (2.600,0.917) -- (2.625,0.935) -- (2.675,0.967) -- (2.700,0.980) -- (2.725,0.985) -- (2.750,0.994) -- (2.775,1.021) -- (2.800,1.034) -- (2.825,1.043) -- (2.850,1.057) -- (2.875,1.075) -- (2.900,1.088) -- (2.925,1.093) -- (2.950,1.106) -- (2.975,1.120) -- (3.000,1.133) -- (3.025,1.169) -- (3.050,1.183) -- (3.075,1.192) -- (3.100,1.206) -- (3.125,1.233) -- (3.150,1.246) -- (3.175,1.278) -- (3.200,1.314) -- (3.225,1.332) -- (3.250,1.354) -- (3.275,1.386) -- (3.300,1.413) -- (3.325,1.426) -- (3.350,1.435) -- (3.375,1.467) -- (3.425,1.494) -- (3.450,1.503) -- (3.475,1.516) -- (3.500,1.539) -- (3.525,1.562) -- (3.550,1.575) -- (3.575,1.589) -- (3.600,1.616) -- (3.625,1.647) -- (3.650,1.652) -- (3.675,1.665) -- (3.700,1.683) -- (3.725,1.692) -- (3.750,1.715) -- (3.775,1.728) -- (3.800,1.746) -- (3.825,1.760) -- (3.850,1.778) -- (3.875,1.809) -- (3.900,1.827) -- (3.925,1.850) -- (3.975,1.877) -- (4.000,1.881) -- (4.025,1.891) -- (4.050,1.909) -- (4.075,1.918) -- (4.100,1.945) -- (4.125,1.990) -- (4.150,2.017) -- (4.175,2.026) -- (4.200,2.039) -- (4.225,2.053) -- (4.250,2.071) -- (4.275,2.093) -- (4.300,2.111) -- (4.325,2.125) -- (4.350,2.156) -- (4.375,2.174) -- (4.400,2.206) -- (4.425,2.233) -- (4.450,2.242) -- (4.475,2.260) -- (4.500,2.283) -- (4.525,2.310) -- (4.550,2.323) -- (4.575,2.359) -- (4.600,2.386) -- (4.625,2.409) -- (4.650,2.418) -- (4.675,2.436) -- (4.700,2.485) -- (4.725,2.503) -- (4.750,2.530) -- (4.775,2.575) -- (4.800,2.603) -- (4.825,2.634) -- (4.850,2.670) -- (4.875,2.697) -- (4.900,2.720) -- (4.925,2.729) -- (4.950,2.747) -- (4.975,2.774) -- (5.000,2.801) -- (5.025,2.837) -- (5.050,2.841) -- (5.075,2.855) -- (5.100,2.873) -- (5.125,2.886) -- (5.150,2.913) -- (5.175,2.918) -- (5.200,2.950) -- (5.225,2.977) -- (5.250,2.995) -- (5.275,3.008) -- (5.300,3.022) -- (5.325,3.040) -- (5.350,3.049) -- (5.375,3.062) -- (5.400,3.071) -- (5.425,3.085) -- (5.450,3.116) -- (5.475,3.125) -- (5.500,3.161) -- (5.500,3.161);
  \draw[cv, cacD, dashed] (0.000,-0.128) -- (0.075,-0.047) -- (0.150,0.034) -- (0.200,0.047) -- (0.250,0.088) -- (0.300,0.128) -- (0.325,0.142) -- (0.375,0.151) -- (0.425,0.169) -- (0.450,0.205) -- (0.500,0.228) -- (0.525,0.232) -- (0.550,0.241) -- (0.575,0.264) -- (0.625,0.273) -- (0.650,0.282) -- (0.675,0.309) -- (0.700,0.340) -- (0.725,0.358) -- (0.750,0.381) -- (0.775,0.403) -- (0.800,0.430) -- (0.825,0.444) -- (0.875,0.457) -- (0.900,0.480) -- (0.925,0.493) -- (0.950,0.511) -- (0.975,0.516) -- (1.000,0.525) -- (1.025,0.543) -- (1.050,0.588) -- (1.075,0.593) -- (1.100,0.606) -- (1.125,0.615) -- (1.150,0.629) -- (1.175,0.633) -- (1.200,0.656) -- (1.225,0.660) -- (1.250,0.669) -- (1.275,0.683) -- (1.300,0.696) -- (1.325,0.705) -- (1.350,0.710) -- (1.375,0.714) -- (1.400,0.719) -- (1.425,0.728) -- (1.450,0.732) -- (1.475,0.750) -- (1.500,0.759) -- (1.525,0.764) -- (1.550,0.777) -- (1.575,0.786) -- (1.600,0.795) -- (1.625,0.813) -- (1.650,0.818) -- (1.675,0.836) -- (1.700,0.840) -- (1.725,0.872) -- (1.750,0.877) -- (1.800,0.886) -- (1.825,0.895) -- (1.850,0.899) -- (1.875,0.908) -- (1.900,0.913) -- (1.950,0.917) -- (1.975,0.926) -- (2.000,0.931) -- (2.025,0.940) -- (2.050,0.953) -- (2.075,0.962) -- (2.100,0.967) -- (2.125,0.976) -- (2.150,0.980) -- (2.175,0.985) -- (2.225,0.989) -- (2.250,0.994) -- (2.275,1.030) -- (2.300,1.043) -- (2.325,1.052) -- (2.350,1.057) -- (2.375,1.070) -- (2.400,1.084) -- (2.425,1.111) -- (2.450,1.124) -- (2.475,1.138) -- (2.500,1.156) -- (2.525,1.183) -- (2.550,1.196) -- (2.575,1.210) -- (2.600,1.224) -- (2.625,1.237) -- (2.650,1.260) -- (2.675,1.273) -- (2.700,1.282) -- (2.725,1.300) -- (2.750,1.327) -- (2.775,1.345) -- (2.800,1.354) -- (2.825,1.377) -- (2.850,1.386) -- (2.875,1.399) -- (2.900,1.426) -- (2.925,1.435) -- (2.950,1.449) -- (2.975,1.453) -- (3.000,1.489) -- (3.025,1.494) -- (3.050,1.512) -- (3.075,1.525) -- (3.100,1.543) -- (3.125,1.571) -- (3.150,1.589) -- (3.175,1.616) -- (3.200,1.629) -- (3.225,1.661) -- (3.250,1.692) -- (3.275,1.706) -- (3.300,1.719) -- (3.325,1.746) -- (3.350,1.755) -- (3.375,1.782) -- (3.400,1.796) -- (3.425,1.827) -- (3.450,1.845) -- (3.475,1.863) -- (3.500,1.881) -- (3.525,1.900) -- (3.550,1.918) -- (3.575,1.945) -- (3.600,1.954) -- (3.625,1.972) -- (3.650,1.976) -- (3.675,1.994) -- (3.700,2.021) -- (3.725,2.035) -- (3.750,2.071) -- (3.775,2.093) -- (3.800,2.111) -- (3.825,2.129) -- (3.850,2.143) -- (3.875,2.156) -- (3.900,2.174) -- (3.925,2.188) -- (3.950,2.192) -- (3.975,2.215) -- (4.000,2.228) -- (4.025,2.242) -- (4.050,2.260) -- (4.075,2.283) -- (4.100,2.287) -- (4.125,2.319) -- (4.150,2.323) -- (4.175,2.346) -- (4.200,2.373) -- (4.225,2.391) -- (4.250,2.400) -- (4.275,2.413) -- (4.300,2.431) -- (4.325,2.445) -- (4.350,2.485) -- (4.375,2.503) -- (4.400,2.517) -- (4.425,2.539) -- (4.450,2.544) -- (4.475,2.557) -- (4.500,2.566) -- (4.525,2.571) -- (4.550,2.585) -- (4.575,2.607) -- (4.600,2.639) -- (4.625,2.652) -- (4.650,2.670) -- (4.675,2.675) -- (4.700,2.688) -- (4.725,2.720) -- (4.750,2.724) -- (4.775,2.729) -- (4.800,2.742) -- (4.825,2.756) -- (4.850,2.774) -- (4.900,2.787) -- (4.925,2.810) -- (4.950,2.832) -- (4.975,2.837) -- (5.000,2.850) -- (5.025,2.868) -- (5.050,2.886) -- (5.075,2.909) -- (5.100,2.927) -- (5.125,2.932) -- (5.150,2.950) -- (5.175,2.963) -- (5.200,2.968) -- (5.225,2.977) -- (5.250,2.981) -- (5.275,2.990) -- (5.300,2.999) -- (5.325,3.008) -- (5.350,3.013) -- (5.375,3.026) -- (5.400,3.044) -- (5.425,3.053) -- (5.450,3.080) -- (5.475,3.085) -- (5.500,3.094) -- (5.500,3.094);
  \draw[cv, cacB, densely dotted] (0.000,-1.269) -- (0.075,-1.201) -- (0.150,-1.142) -- (0.200,-1.129) -- (0.250,-1.102) -- (0.300,-1.057) -- (0.325,-1.021) -- (0.375,-0.967) -- (0.425,-0.922) -- (0.450,-0.886) -- (0.500,-0.863) -- (0.525,-0.854) -- (0.550,-0.813) -- (0.575,-0.773) -- (0.625,-0.741) -- (0.650,-0.732) -- (0.675,-0.719) -- (0.700,-0.687) -- (0.725,-0.656) -- (0.750,-0.638) -- (0.775,-0.611) -- (0.800,-0.552) -- (0.825,-0.530) -- (0.850,-0.511) -- (0.875,-0.484) -- (0.900,-0.448) -- (0.925,-0.439) -- (0.950,-0.417) -- (0.975,-0.390) -- (1.000,-0.381) -- (1.025,-0.349) -- (1.050,-0.327) -- (1.075,-0.291) -- (1.100,-0.273) -- (1.125,-0.223) -- (1.150,-0.214) -- (1.175,-0.210) -- (1.200,-0.187) -- (1.225,-0.151) -- (1.250,-0.137) -- (1.275,-0.115) -- (1.300,-0.083) -- (1.325,-0.052) -- (1.350,-0.025) -- (1.375,-0.011) -- (1.400,0.002) -- (1.425,0.034) -- (1.450,0.056) -- (1.475,0.079) -- (1.500,0.110) -- (1.525,0.137) -- (1.550,0.146) -- (1.575,0.183) -- (1.600,0.192) -- (1.625,0.219) -- (1.650,0.228) -- (1.675,0.241) -- (1.700,0.264) -- (1.725,0.277) -- (1.750,0.300) -- (1.775,0.318) -- (1.800,0.340) -- (1.825,0.363) -- (1.850,0.372) -- (1.875,0.394) -- (1.900,0.412) -- (1.925,0.417) -- (1.950,0.430) -- (1.975,0.444) -- (2.000,0.448) -- (2.025,0.457) -- (2.050,0.475) -- (2.075,0.489) -- (2.100,0.502) -- (2.125,0.530) -- (2.150,0.539) -- (2.175,0.543) -- (2.200,0.548) -- (2.225,0.579) -- (2.250,0.588) -- (2.275,0.611) -- (2.300,0.629) -- (2.325,0.647) -- (2.350,0.651) -- (2.375,0.665) -- (2.400,0.696) -- (2.425,0.714) -- (2.450,0.728) -- (2.475,0.746) -- (2.500,0.750) -- (2.525,0.764) -- (2.550,0.786) -- (2.575,0.804) -- (2.600,0.818) -- (2.650,0.836) -- (2.675,0.858) -- (2.700,0.872) -- (2.725,0.881) -- (2.750,0.886) -- (2.775,0.895) -- (2.800,0.899) -- (2.825,0.908) -- (2.850,0.926) -- (2.900,0.931) -- (2.925,0.958) -- (2.950,0.980) -- (2.975,1.007) -- (3.000,1.043) -- (3.025,1.052) -- (3.050,1.075) -- (3.075,1.124) -- (3.100,1.156) -- (3.125,1.174) -- (3.150,1.192) -- (3.175,1.206) -- (3.200,1.237) -- (3.225,1.264) -- (3.250,1.273) -- (3.275,1.278) -- (3.300,1.291) -- (3.325,1.296) -- (3.350,1.323) -- (3.375,1.341) -- (3.400,1.377) -- (3.425,1.404) -- (3.450,1.417) -- (3.475,1.422) -- (3.500,1.444) -- (3.525,1.453) -- (3.550,1.471) -- (3.575,1.480) -- (3.600,1.485) -- (3.625,1.498) -- (3.650,1.516) -- (3.675,1.530) -- (3.700,1.553) -- (3.725,1.557) -- (3.750,1.575) -- (3.775,1.598) -- (3.800,1.616) -- (3.850,1.625) -- (3.875,1.643) -- (3.900,1.665) -- (3.925,1.674) -- (3.950,1.692) -- (3.975,1.701) -- (4.000,1.710) -- (4.025,1.751) -- (4.050,1.778) -- (4.075,1.796) -- (4.100,1.805) -- (4.125,1.823) -- (4.150,1.841) -- (4.175,1.850) -- (4.200,1.863) -- (4.225,1.900) -- (4.250,1.927) -- (4.275,1.945) -- (4.300,1.967) -- (4.325,1.981) -- (4.350,1.999) -- (4.375,2.017) -- (4.400,2.057) -- (4.425,2.089) -- (4.450,2.102) -- (4.475,2.111) -- (4.500,2.125) -- (4.525,2.147) -- (4.550,2.170) -- (4.575,2.192) -- (4.600,2.219) -- (4.625,2.251) -- (4.650,2.269) -- (4.675,2.301) -- (4.700,2.310) -- (4.725,2.346) -- (4.750,2.373) -- (4.775,2.377) -- (4.800,2.409) -- (4.825,2.445) -- (4.850,2.485) -- (4.875,2.499) -- (4.900,2.521) -- (4.925,2.539) -- (4.950,2.557) -- (4.975,2.580) -- (5.000,2.594) -- (5.025,2.607) -- (5.050,2.625) -- (5.075,2.648) -- (5.100,2.675) -- (5.125,2.697) -- (5.150,2.724) -- (5.175,2.760) -- (5.200,2.778) -- (5.225,2.805) -- (5.250,2.819) -- (5.275,2.828) -- (5.300,2.837) -- (5.325,2.855) -- (5.350,2.868) -- (5.375,2.886) -- (5.400,2.895) -- (5.425,2.923) -- (5.450,2.945) -- (5.475,2.954) -- (5.500,2.972) -- (5.500,2.972);
  \draw[cv, cacC, solid] (0.000,-1.251) -- (0.075,-1.192) -- (0.150,-1.138) -- (0.200,-1.115) -- (0.250,-1.061) -- (0.300,-1.016) -- (0.325,-0.976) -- (0.375,-0.940) -- (0.425,-0.908) -- (0.450,-0.845) -- (0.500,-0.813) -- (0.525,-0.791) -- (0.550,-0.786) -- (0.575,-0.750) -- (0.625,-0.741) -- (0.650,-0.723) -- (0.675,-0.710) -- (0.700,-0.669) -- (0.725,-0.647) -- (0.750,-0.638) -- (0.775,-0.620) -- (0.800,-0.579) -- (0.825,-0.570) -- (0.850,-0.534) -- (0.875,-0.525) -- (0.900,-0.480) -- (0.925,-0.457) -- (0.950,-0.439) -- (0.975,-0.421) -- (1.000,-0.403) -- (1.025,-0.372) -- (1.050,-0.340) -- (1.075,-0.327) -- (1.100,-0.304) -- (1.125,-0.255) -- (1.150,-0.237) -- (1.175,-0.219) -- (1.200,-0.201) -- (1.225,-0.160) -- (1.250,-0.133) -- (1.275,-0.106) -- (1.300,-0.088) -- (1.325,-0.065) -- (1.350,-0.038) -- (1.375,-0.020) -- (1.400,0.002) -- (1.425,0.034) -- (1.450,0.052) -- (1.475,0.065) -- (1.500,0.101) -- (1.525,0.155) -- (1.550,0.174) -- (1.575,0.201) -- (1.600,0.205) -- (1.625,0.210) -- (1.650,0.214) -- (1.675,0.219) -- (1.700,0.232) -- (1.725,0.250) -- (1.750,0.277) -- (1.775,0.304) -- (1.800,0.318) -- (1.825,0.327) -- (1.850,0.345) -- (1.875,0.372) -- (1.900,0.381) -- (1.925,0.399) -- (1.950,0.408) -- (1.975,0.435) -- (2.000,0.448) -- (2.025,0.457) -- (2.050,0.480) -- (2.075,0.493) -- (2.100,0.498) -- (2.125,0.511) -- (2.150,0.530) -- (2.175,0.543) -- (2.200,0.561) -- (2.225,0.575) -- (2.250,0.593) -- (2.275,0.606) -- (2.300,0.611) -- (2.325,0.629) -- (2.350,0.642) -- (2.375,0.656) -- (2.400,0.674) -- (2.425,0.683) -- (2.450,0.723) -- (2.475,0.737) -- (2.500,0.755) -- (2.525,0.773) -- (2.550,0.795) -- (2.575,0.813) -- (2.600,0.836) -- (2.625,0.840) -- (2.650,0.868) -- (2.675,0.877) -- (2.700,0.890) -- (2.725,0.899) -- (2.750,0.931) -- (2.775,0.944) -- (2.800,0.967) -- (2.825,0.980) -- (2.850,0.998) -- (2.875,1.016) -- (2.900,1.039) -- (2.925,1.061) -- (2.950,1.079) -- (2.975,1.106) -- (3.000,1.133) -- (3.025,1.160) -- (3.050,1.169) -- (3.075,1.196) -- (3.100,1.215) -- (3.125,1.251) -- (3.150,1.278) -- (3.175,1.296) -- (3.200,1.314) -- (3.225,1.341) -- (3.250,1.377) -- (3.275,1.395) -- (3.300,1.408) -- (3.325,1.426) -- (3.350,1.449) -- (3.375,1.476) -- (3.400,1.498) -- (3.425,1.516) -- (3.450,1.539) -- (3.475,1.553) -- (3.500,1.580) -- (3.525,1.602) -- (3.550,1.620) -- (3.575,1.643) -- (3.600,1.661) -- (3.625,1.679) -- (3.650,1.715) -- (3.675,1.737) -- (3.700,1.755) -- (3.725,1.778) -- (3.750,1.800) -- (3.775,1.841) -- (3.800,1.863) -- (3.825,1.891) -- (3.850,1.931) -- (3.875,1.945) -- (3.900,1.958) -- (3.925,1.990) -- (3.950,2.008) -- (3.975,2.035) -- (4.000,2.048) -- (4.025,2.057) -- (4.050,2.066) -- (4.075,2.084) -- (4.100,2.107) -- (4.125,2.120) -- (4.150,2.138) -- (4.175,2.161) -- (4.200,2.188) -- (4.225,2.210) -- (4.250,2.228) -- (4.275,2.251) -- (4.300,2.269) -- (4.325,2.305) -- (4.350,2.328) -- (4.375,2.341) -- (4.400,2.355) -- (4.425,2.391) -- (4.450,2.418) -- (4.500,2.436) -- (4.525,2.463) -- (4.550,2.490) -- (4.575,2.526) -- (4.600,2.557) -- (4.625,2.603) -- (4.650,2.625) -- (4.675,2.652) -- (4.700,2.666) -- (4.725,2.697) -- (4.750,2.724) -- (4.775,2.729) -- (4.800,2.733) -- (4.825,2.751) -- (4.850,2.778) -- (4.875,2.792) -- (4.900,2.810) -- (4.925,2.832) -- (4.950,2.859) -- (4.975,2.873) -- (5.000,2.900) -- (5.025,2.923) -- (5.050,2.945) -- (5.075,2.959) -- (5.100,2.977) -- (5.125,3.004) -- (5.150,3.026) -- (5.175,3.049) -- (5.200,3.062) -- (5.225,3.085) -- (5.250,3.089) -- (5.275,3.112) -- (5.300,3.139) -- (5.325,3.152) -- (5.350,3.170) -- (5.375,3.179) -- (5.400,3.188) -- (5.425,3.197) -- (5.450,3.224) -- (5.500,3.256) -- (5.500,3.256);
  \end{scope}
  \node[font=\sffamily\small\bfseries] at (9.87,4.37) {SMT-COMP selection, \num{1200}\,s};
  \draw[gr] (7.120,0) -- (7.120,3.95);
  \draw[gm] (7.446,0) -- (7.446,3.95);
  \draw[gm] (7.637,0) -- (7.637,3.95);
  \draw[gm] (7.772,0) -- (7.772,3.95);
  \draw[gm] (7.877,0) -- (7.877,3.95);
  \draw[gm] (7.963,0) -- (7.963,3.95);
  \draw[gm] (8.035,0) -- (8.035,3.95);
  \draw[gm] (8.098,0) -- (8.098,3.95);
  \draw[gm] (8.153,0) -- (8.153,3.95);
  \draw[gr] (8.203,0) -- (8.203,3.95);
  \draw[gm] (8.529,0) -- (8.529,3.95);
  \draw[gm] (8.720,0) -- (8.720,3.95);
  \draw[gm] (8.855,0) -- (8.855,3.95);
  \draw[gm] (8.960,0) -- (8.960,3.95);
  \draw[gm] (9.045,0) -- (9.045,3.95);
  \draw[gm] (9.118,0) -- (9.118,3.95);
  \draw[gm] (9.181,0) -- (9.181,3.95);
  \draw[gm] (9.236,0) -- (9.236,3.95);
  \draw[gr] (9.286,0) -- (9.286,3.95);
  \draw[gm] (9.612,0) -- (9.612,3.95);
  \draw[gm] (9.802,0) -- (9.802,3.95);
  \draw[gm] (9.938,0) -- (9.938,3.95);
  \draw[gm] (10.043,0) -- (10.043,3.95);
  \draw[gm] (10.128,0) -- (10.128,3.95);
  \draw[gm] (10.201,0) -- (10.201,3.95);
  \draw[gm] (10.264,0) -- (10.264,3.95);
  \draw[gm] (10.319,0) -- (10.319,3.95);
  \draw[gr] (10.369,0) -- (10.369,3.95);
  \draw[gm] (10.695,0) -- (10.695,3.95);
  \draw[gm] (10.885,0) -- (10.885,3.95);
  \draw[gm] (11.020,0) -- (11.020,3.95);
  \draw[gm] (11.125,0) -- (11.125,3.95);
  \draw[gm] (11.211,0) -- (11.211,3.95);
  \draw[gm] (11.284,0) -- (11.284,3.95);
  \draw[gm] (11.346,0) -- (11.346,3.95);
  \draw[gm] (11.402,0) -- (11.402,3.95);
  \draw[gr] (11.451,0) -- (11.451,3.95);
  \draw[gm] (11.777,0) -- (11.777,3.95);
  \draw[gm] (11.968,0) -- (11.968,3.95);
  \draw[gm] (12.103,0) -- (12.103,3.95);
  \draw[gm] (12.208,0) -- (12.208,3.95);
  \draw[gm] (12.294,0) -- (12.294,3.95);
  \draw[gm] (12.367,0) -- (12.367,3.95);
  \draw[gm] (12.429,0) -- (12.429,3.95);
  \draw[gm] (12.485,0) -- (12.485,3.95);
  \draw[gr] (12.534,0) -- (12.534,3.95);
  \draw[gr] (7.12,0.000) -- (12.62,0.000);
  \node[anchor=east] at (7.04,0.000) {20};
  \draw[gr] (7.12,0.988) -- (12.62,0.988);
  \node[anchor=east] at (7.04,0.988) {40};
  \draw[gr] (7.12,1.975) -- (12.62,1.975);
  \node[anchor=east] at (7.04,1.975) {60};
  \draw[gr] (7.12,2.963) -- (12.62,2.963);
  \node[anchor=east] at (7.04,2.963) {80};
  \draw[gr] (7.12,3.950) -- (12.62,3.950);
  \node[anchor=east] at (7.04,3.950) {100};
  \draw[ax] (7.12,0) -- (7.12,-0.09);
  \node[anchor=north] at (7.12,-0.12) {0.01};
  \draw[ax] (8.20,0) -- (8.20,-0.09);
  \node[anchor=north] at (8.20,-0.12) {0.1};
  \draw[ax] (9.29,0) -- (9.29,-0.09);
  \node[anchor=north] at (9.29,-0.12) {1};
  \draw[ax] (10.37,0) -- (10.37,-0.09);
  \node[anchor=north] at (10.37,-0.12) {10};
  \draw[ax] (11.45,0) -- (11.45,-0.09);
  \node[anchor=north] at (11.45,-0.12) {100};
  \draw[ax] (12.53,0) -- (12.53,-0.09);
  \node[anchor=north] at (12.53,-0.12) {1000};
  \draw[ax] (7.12,0) -- (12.62,0);
  \draw[ax] (7.12,0) -- (7.12,3.95);
  \node[anchor=north] at (9.87,-0.52) {time (s)};
  \node[rotate=90, anchor=south] at (6.24,1.98) {solved [\%]};
  \begin{scope}
  \clip (7.12,-0.02) rectangle (12.62,4.00);
  \draw[cv, cacA, densely dashed] (7.120,-0.493) -- (7.170,-0.436) -- (7.220,-0.379) -- (7.245,-0.341) -- (7.295,-0.284) -- (7.320,-0.236) -- (7.345,-0.188) -- (7.370,-0.179) -- (7.420,-0.169) -- (7.445,-0.122) -- (7.470,-0.103) -- (7.495,-0.074) -- (7.520,-0.046) -- (7.545,-0.017) -- (7.570,0.002) -- (7.595,0.040) -- (7.620,0.059) -- (7.670,0.078) -- (7.695,0.097) -- (7.720,0.116) -- (7.745,0.135) -- (7.770,0.164) -- (7.795,0.202) -- (7.820,0.221) -- (7.895,0.240) -- (7.920,0.249) -- (7.970,0.278) -- (7.995,0.316) -- (8.020,0.335) -- (8.045,0.373) -- (8.070,0.392) -- (8.095,0.420) -- (8.120,0.440) -- (8.145,0.468) -- (8.170,0.478) -- (8.195,0.525) -- (8.220,0.544) -- (8.245,0.554) -- (8.270,0.573) -- (8.295,0.601) -- (8.320,0.611) -- (8.345,0.630) -- (8.370,0.658) -- (8.395,0.677) -- (8.420,0.696) -- (8.445,0.715) -- (8.470,0.753) -- (8.495,0.772) -- (8.520,0.792) -- (8.545,0.820) -- (8.570,0.868) -- (8.595,0.896) -- (8.620,0.906) -- (8.645,0.925) -- (8.670,0.953) -- (8.695,0.963) -- (8.720,1.001) -- (8.745,1.048) -- (8.770,1.058) -- (8.795,1.086) -- (8.820,1.105) -- (8.845,1.115) -- (8.870,1.134) -- (8.895,1.144) -- (8.920,1.163) -- (8.945,1.191) -- (8.970,1.210) -- (8.995,1.220) -- (9.020,1.229) -- (9.045,1.239) -- (9.070,1.258) -- (9.095,1.277) -- (9.120,1.296) -- (9.145,1.315) -- (9.170,1.353) -- (9.195,1.381) -- (9.220,1.400) -- (9.245,1.419) -- (9.270,1.457) -- (9.295,1.496) -- (9.320,1.515) -- (9.345,1.524) -- (9.370,1.591) -- (9.420,1.648) -- (9.445,1.657) -- (9.470,1.686) -- (9.495,1.705) -- (9.520,1.724) -- (9.545,1.762) -- (9.570,1.771) -- (9.595,1.809) -- (9.620,1.838) -- (9.645,1.848) -- (9.670,1.857) -- (9.695,1.867) -- (9.745,1.886) -- (9.770,1.952) -- (9.795,1.971) -- (9.820,2.000) -- (9.845,2.047) -- (9.895,2.085) -- (9.920,2.123) -- (9.945,2.152) -- (9.970,2.200) -- (9.995,2.209) -- (10.020,2.219) -- (10.045,2.266) -- (10.070,2.314) -- (10.095,2.333) -- (10.120,2.361) -- (10.145,2.390) -- (10.170,2.409) -- (10.195,2.428) -- (10.220,2.466) -- (10.245,2.504) -- (10.270,2.542) -- (10.295,2.571) -- (10.320,2.590) -- (10.345,2.628) -- (10.370,2.656) -- (10.395,2.694) -- (10.420,2.704) -- (10.445,2.770) -- (10.470,2.875) -- (10.495,2.894) -- (10.520,2.913) -- (10.545,2.951) -- (10.570,2.989) -- (10.595,3.008) -- (10.620,3.037) -- (10.645,3.065) -- (10.670,3.084) -- (10.695,3.094) -- (10.720,3.113) -- (10.745,3.141) -- (10.770,3.160) -- (10.795,3.179) -- (10.820,3.189) -- (10.845,3.217) -- (10.870,3.227) -- (10.895,3.256) -- (10.920,3.284) -- (10.945,3.294) -- (10.970,3.303) -- (11.020,3.313) -- (11.045,3.332) -- (11.070,3.341) -- (11.120,3.360) -- (11.145,3.370) -- (11.170,3.379) -- (11.195,3.408) -- (11.220,3.417) -- (11.245,3.436) -- (11.270,3.446) -- (11.320,3.474) -- (11.345,3.484) -- (11.495,3.503) -- (11.520,3.512) -- (11.545,3.531) -- (11.595,3.541) -- (11.645,3.550) -- (11.670,3.569) -- (11.695,3.579) -- (11.720,3.608) -- (11.745,3.617) -- (11.845,3.636) -- (11.945,3.646) -- (11.970,3.655) -- (11.995,3.674) -- (12.020,3.684) -- (12.070,3.693) -- (12.170,3.703) -- (12.195,3.712) -- (12.220,3.722) -- (12.245,3.731) -- (12.270,3.741) -- (12.345,3.750) -- (12.395,3.760) -- (12.445,3.779) -- (12.595,3.788) -- (12.620,3.788);
  \draw[cv, cacD, dashed] (7.120,0.240) -- (7.170,0.287) -- (7.220,0.335) -- (7.245,0.363) -- (7.295,0.392) -- (7.320,0.411) -- (7.345,0.430) -- (7.370,0.449) -- (7.445,0.506) -- (7.470,0.525) -- (7.495,0.544) -- (7.520,0.563) -- (7.545,0.592) -- (7.570,0.630) -- (7.595,0.658) -- (7.645,0.677) -- (7.670,0.696) -- (7.695,0.706) -- (7.720,0.725) -- (7.745,0.734) -- (7.770,0.763) -- (7.795,0.792) -- (7.820,0.820) -- (7.845,0.830) -- (7.870,0.839) -- (7.920,0.868) -- (7.945,0.877) -- (7.970,0.887) -- (7.995,0.896) -- (8.020,0.906) -- (8.045,0.915) -- (8.070,0.925) -- (8.095,0.963) -- (8.120,0.972) -- (8.195,0.991) -- (8.220,1.001) -- (8.245,1.029) -- (8.270,1.048) -- (8.320,1.058) -- (8.345,1.067) -- (8.370,1.077) -- (8.470,1.096) -- (8.495,1.105) -- (8.520,1.134) -- (8.595,1.144) -- (8.645,1.163) -- (8.695,1.210) -- (8.720,1.239) -- (8.745,1.248) -- (8.770,1.258) -- (8.795,1.296) -- (8.820,1.324) -- (8.845,1.353) -- (8.870,1.372) -- (8.895,1.381) -- (8.920,1.400) -- (8.945,1.429) -- (8.970,1.438) -- (8.995,1.457) -- (9.020,1.476) -- (9.045,1.496) -- (9.070,1.524) -- (9.095,1.562) -- (9.120,1.581) -- (9.145,1.610) -- (9.170,1.657) -- (9.195,1.686) -- (9.220,1.714) -- (9.245,1.752) -- (9.270,1.790) -- (9.295,1.838) -- (9.320,1.867) -- (9.345,1.895) -- (9.370,1.905) -- (9.395,1.924) -- (9.420,1.943) -- (9.445,1.981) -- (9.470,2.019) -- (9.495,2.047) -- (9.520,2.057) -- (9.545,2.066) -- (9.570,2.085) -- (9.595,2.095) -- (9.645,2.123) -- (9.670,2.180) -- (9.695,2.219) -- (9.720,2.247) -- (9.745,2.276) -- (9.770,2.285) -- (9.795,2.323) -- (9.820,2.342) -- (9.845,2.380) -- (9.895,2.418) -- (9.920,2.456) -- (9.945,2.485) -- (9.970,2.513) -- (9.995,2.542) -- (10.070,2.571) -- (10.095,2.609) -- (10.120,2.618) -- (10.145,2.637) -- (10.170,2.647) -- (10.195,2.666) -- (10.220,2.694) -- (10.270,2.732) -- (10.295,2.761) -- (10.345,2.780) -- (10.370,2.799) -- (10.420,2.808) -- (10.445,2.818) -- (10.470,2.846) -- (10.495,2.913) -- (10.545,2.932) -- (10.570,2.942) -- (10.595,2.989) -- (10.620,2.999) -- (10.645,3.018) -- (10.670,3.046) -- (10.695,3.056) -- (10.720,3.065) -- (10.795,3.103) -- (10.820,3.132) -- (10.845,3.141) -- (10.870,3.160) -- (10.920,3.179) -- (10.945,3.189) -- (10.970,3.208) -- (11.020,3.227) -- (11.045,3.236) -- (11.070,3.256) -- (11.095,3.265) -- (11.120,3.275) -- (11.195,3.294) -- (11.220,3.303) -- (11.245,3.313) -- (11.295,3.322) -- (11.470,3.341) -- (11.495,3.370) -- (11.545,3.398) -- (11.570,3.436) -- (11.620,3.465) -- (11.645,3.484) -- (11.695,3.503) -- (11.720,3.512) -- (11.745,3.522) -- (11.770,3.531) -- (11.845,3.541) -- (11.870,3.550) -- (11.945,3.569) -- (12.020,3.579) -- (12.070,3.608) -- (12.095,3.617) -- (12.170,3.636) -- (12.195,3.646) -- (12.245,3.665) -- (12.270,3.674) -- (12.395,3.684) -- (12.495,3.693) -- (12.570,3.712) -- (12.620,3.722) -- (12.620,3.722);
  \draw[cv, cacB, densely dotted] (7.120,-0.816) -- (7.170,-0.769) -- (7.220,-0.683) -- (7.245,-0.655) -- (7.295,-0.645) -- (7.320,-0.597) -- (7.345,-0.578) -- (7.370,-0.559) -- (7.420,-0.531) -- (7.445,-0.502) -- (7.470,-0.464) -- (7.495,-0.455) -- (7.520,-0.417) -- (7.570,-0.331) -- (7.595,-0.284) -- (7.620,-0.264) -- (7.645,-0.198) -- (7.670,-0.160) -- (7.695,-0.150) -- (7.720,-0.112) -- (7.745,-0.103) -- (7.770,-0.046) -- (7.795,-0.017) -- (7.820,0.011) -- (7.845,0.059) -- (7.870,0.097) -- (7.895,0.126) -- (7.920,0.145) -- (7.945,0.183) -- (7.970,0.221) -- (7.995,0.268) -- (8.020,0.287) -- (8.045,0.316) -- (8.070,0.344) -- (8.095,0.354) -- (8.120,0.382) -- (8.145,0.430) -- (8.195,0.497) -- (8.220,0.525) -- (8.245,0.544) -- (8.270,0.563) -- (8.295,0.601) -- (8.320,0.620) -- (8.345,0.658) -- (8.370,0.668) -- (8.395,0.687) -- (8.420,0.706) -- (8.445,0.715) -- (8.470,0.725) -- (8.495,0.734) -- (8.520,0.753) -- (8.545,0.772) -- (8.570,0.801) -- (8.595,0.820) -- (8.620,0.830) -- (8.645,0.839) -- (8.670,0.877) -- (8.695,0.896) -- (8.720,0.906) -- (8.745,0.934) -- (8.770,0.972) -- (8.795,1.001) -- (8.820,1.020) -- (8.870,1.077) -- (8.895,1.096) -- (8.920,1.105) -- (8.945,1.144) -- (8.995,1.163) -- (9.070,1.182) -- (9.095,1.201) -- (9.120,1.220) -- (9.145,1.267) -- (9.170,1.296) -- (9.195,1.324) -- (9.220,1.362) -- (9.245,1.381) -- (9.270,1.429) -- (9.295,1.438) -- (9.320,1.476) -- (9.345,1.505) -- (9.420,1.515) -- (9.445,1.524) -- (9.470,1.534) -- (9.495,1.543) -- (9.545,1.572) -- (9.570,1.591) -- (9.595,1.610) -- (9.645,1.619) -- (9.670,1.629) -- (9.695,1.638) -- (9.720,1.648) -- (9.745,1.657) -- (9.795,1.705) -- (9.820,1.724) -- (9.845,1.752) -- (9.870,1.790) -- (9.895,1.838) -- (9.920,1.857) -- (9.945,1.905) -- (9.970,1.962) -- (9.995,1.981) -- (10.020,2.019) -- (10.045,2.028) -- (10.070,2.066) -- (10.095,2.104) -- (10.120,2.161) -- (10.145,2.200) -- (10.170,2.238) -- (10.195,2.266) -- (10.220,2.285) -- (10.245,2.295) -- (10.270,2.323) -- (10.295,2.333) -- (10.320,2.409) -- (10.345,2.428) -- (10.370,2.466) -- (10.395,2.494) -- (10.420,2.552) -- (10.445,2.561) -- (10.470,2.580) -- (10.495,2.609) -- (10.520,2.628) -- (10.545,2.656) -- (10.570,2.675) -- (10.595,2.694) -- (10.620,2.732) -- (10.645,2.780) -- (10.670,2.837) -- (10.695,2.846) -- (10.720,2.865) -- (10.745,2.923) -- (10.770,2.932) -- (10.795,2.951) -- (10.845,2.989) -- (10.870,3.008) -- (10.895,3.037) -- (10.970,3.084) -- (10.995,3.103) -- (11.020,3.141) -- (11.070,3.170) -- (11.095,3.208) -- (11.120,3.227) -- (11.145,3.256) -- (11.170,3.265) -- (11.220,3.275) -- (11.245,3.303) -- (11.270,3.313) -- (11.295,3.322) -- (11.320,3.341) -- (11.345,3.360) -- (11.370,3.370) -- (11.395,3.379) -- (11.420,3.398) -- (11.445,3.417) -- (11.470,3.436) -- (11.520,3.455) -- (11.545,3.465) -- (11.570,3.474) -- (11.595,3.493) -- (11.670,3.503) -- (11.695,3.512) -- (11.720,3.522) -- (11.795,3.560) -- (11.845,3.579) -- (11.870,3.598) -- (11.920,3.617) -- (11.945,3.627) -- (11.970,3.636) -- (12.020,3.646) -- (12.220,3.655) -- (12.245,3.665) -- (12.270,3.674) -- (12.320,3.693) -- (12.345,3.703) -- (12.420,3.722) -- (12.545,3.731) -- (12.595,3.750) -- (12.620,3.760) -- (12.620,3.760);
  \draw[cv, cacC, solid] (7.120,-0.769) -- (7.170,-0.731) -- (7.220,-0.712) -- (7.245,-0.702) -- (7.295,-0.664) -- (7.320,-0.626) -- (7.345,-0.597) -- (7.370,-0.550) -- (7.420,-0.521) -- (7.445,-0.493) -- (7.470,-0.417) -- (7.495,-0.407) -- (7.520,-0.388) -- (7.545,-0.369) -- (7.570,-0.350) -- (7.595,-0.284) -- (7.620,-0.264) -- (7.645,-0.198) -- (7.670,-0.150) -- (7.695,-0.122) -- (7.720,-0.103) -- (7.745,-0.084) -- (7.795,-0.074) -- (7.820,-0.017) -- (7.845,-0.008) -- (7.870,0.030) -- (7.895,0.040) -- (7.920,0.088) -- (7.945,0.116) -- (7.970,0.145) -- (7.995,0.202) -- (8.020,0.230) -- (8.045,0.287) -- (8.070,0.325) -- (8.095,0.382) -- (8.145,0.420) -- (8.170,0.430) -- (8.195,0.440) -- (8.220,0.468) -- (8.245,0.497) -- (8.270,0.516) -- (8.295,0.544) -- (8.320,0.582) -- (8.345,0.620) -- (8.370,0.630) -- (8.395,0.649) -- (8.420,0.668) -- (8.445,0.696) -- (8.470,0.706) -- (8.520,0.753) -- (8.545,0.772) -- (8.570,0.792) -- (8.595,0.811) -- (8.620,0.830) -- (8.645,0.849) -- (8.670,0.858) -- (8.695,0.896) -- (8.720,0.915) -- (8.745,0.925) -- (8.770,0.972) -- (8.795,0.982) -- (8.820,1.001) -- (8.845,1.020) -- (8.870,1.039) -- (8.895,1.067) -- (8.920,1.086) -- (8.945,1.115) -- (8.970,1.134) -- (8.995,1.144) -- (9.020,1.163) -- (9.045,1.191) -- (9.070,1.201) -- (9.095,1.210) -- (9.120,1.239) -- (9.145,1.296) -- (9.170,1.324) -- (9.195,1.353) -- (9.220,1.372) -- (9.245,1.410) -- (9.270,1.448) -- (9.295,1.515) -- (9.320,1.534) -- (9.345,1.543) -- (9.370,1.562) -- (9.420,1.581) -- (9.445,1.610) -- (9.470,1.676) -- (9.495,1.705) -- (9.520,1.724) -- (9.545,1.743) -- (9.570,1.771) -- (9.595,1.819) -- (9.620,1.876) -- (9.645,1.905) -- (9.670,1.943) -- (9.695,1.952) -- (9.720,1.990) -- (9.745,2.038) -- (9.770,2.076) -- (9.795,2.104) -- (9.820,2.133) -- (9.845,2.152) -- (9.870,2.180) -- (9.895,2.209) -- (9.920,2.219) -- (9.970,2.257) -- (9.995,2.304) -- (10.020,2.333) -- (10.045,2.352) -- (10.070,2.390) -- (10.095,2.418) -- (10.120,2.437) -- (10.145,2.466) -- (10.170,2.475) -- (10.195,2.504) -- (10.220,2.523) -- (10.245,2.571) -- (10.270,2.618) -- (10.295,2.666) -- (10.320,2.704) -- (10.345,2.761) -- (10.370,2.780) -- (10.395,2.808) -- (10.420,2.846) -- (10.445,2.875) -- (10.470,2.884) -- (10.495,2.904) -- (10.520,2.913) -- (10.545,2.961) -- (10.570,2.999) -- (10.595,3.046) -- (10.620,3.065) -- (10.645,3.103) -- (10.670,3.122) -- (10.695,3.141) -- (10.720,3.170) -- (10.745,3.189) -- (10.770,3.217) -- (10.795,3.227) -- (10.820,3.246) -- (10.845,3.265) -- (10.895,3.294) -- (10.920,3.322) -- (10.945,3.351) -- (10.970,3.370) -- (10.995,3.398) -- (11.020,3.417) -- (11.070,3.436) -- (11.095,3.455) -- (11.120,3.474) -- (11.195,3.484) -- (11.245,3.493) -- (11.295,3.503) -- (11.320,3.522) -- (11.370,3.531) -- (11.395,3.541) -- (11.420,3.569) -- (11.445,3.579) -- (11.520,3.588) -- (11.545,3.608) -- (11.595,3.617) -- (11.645,3.627) -- (11.670,3.646) -- (11.720,3.655) -- (11.745,3.665) -- (11.770,3.674) -- (11.820,3.684) -- (11.845,3.693) -- (11.995,3.703) -- (12.045,3.712) -- (12.070,3.731) -- (12.170,3.750) -- (12.270,3.760) -- (12.370,3.769) -- (12.395,3.788) -- (12.445,3.807) -- (12.620,3.807);
  \end{scope}
  \draw[cv, cacA, densely dashed] (2.64,-1.05) -- (3.10,-1.05);
  \node[anchor=west] at (3.20,-1.05) {\opensmt{}};
  \draw[cv, cacD, dashed] (4.99,-1.05) -- (5.45,-1.05);
  \node[anchor=west] at (5.55,-1.05) {\solvername{yices}};
  \draw[cv, cacB, densely dotted] (6.74,-1.05) -- (7.20,-1.05);
  \node[anchor=west] at (7.30,-1.05) {\primo{}};
  \draw[cv, cacC, solid] (8.49,-1.05) -- (8.95,-1.05);
  \node[anchor=west] at (9.05,-1.05) {\primobest{}};
\end{tikzpicture}
\caption{Percentage of instances solved against wall-clock time.}
\label{fig:cactus}
\end{figure}
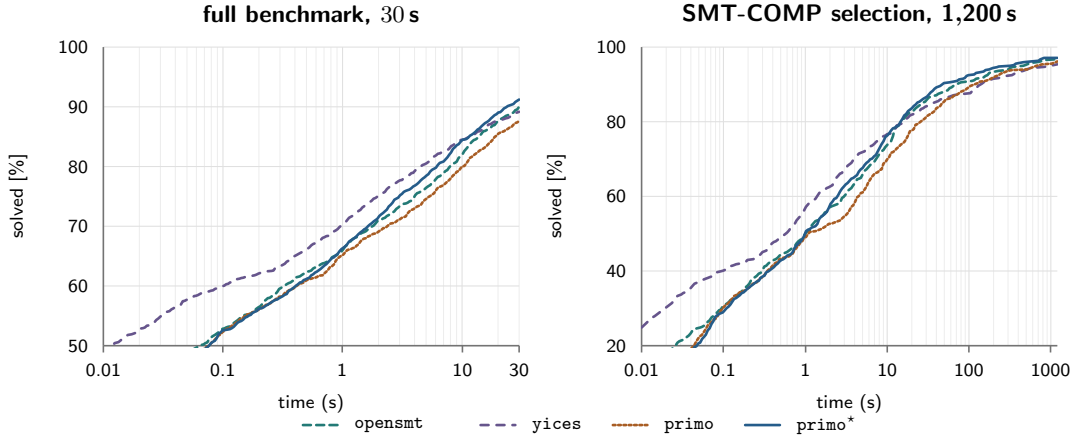

\begin{figure}[H]
  \begin{center}
    \includegraphics[width=0.70\textwidth,trim=35 4 70 9,clip]%
      {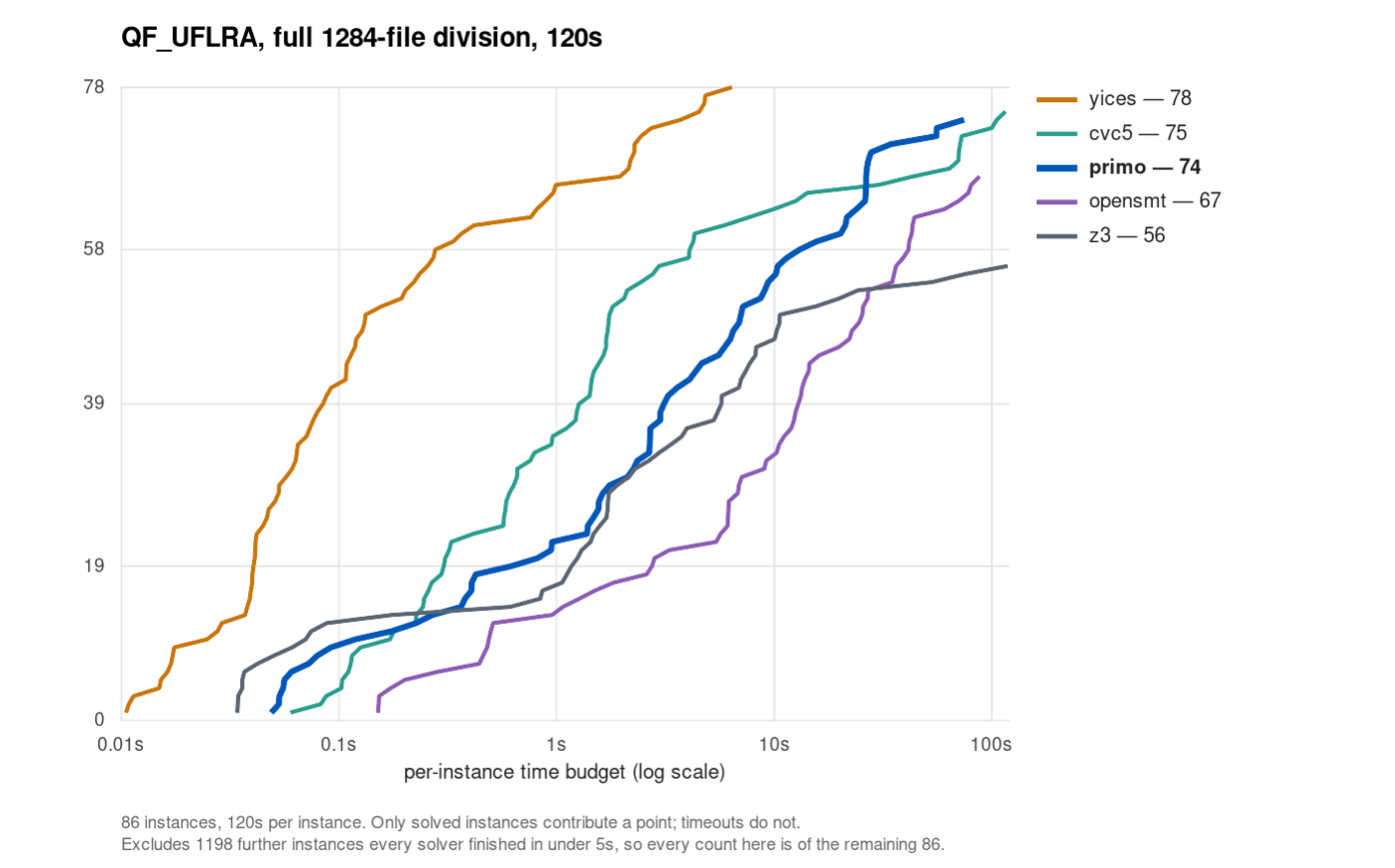}
  \end{center}
  \caption{Comparison of solvers on QF-UFLRA.}\label{fig:uflra}
\end{figure}

\autoref{fig:cactus} plots the two QF-LRA evaluations of
\autoref{sec:experiments}, the full benchmark at $30$\,s and the SMT-COMP
selection at \num{1200}\,s.  \autoref{fig:uflra} compares \primo{} with four
other solvers on QF-UFLRA, a different recently-implemented logic combination,
at a $120$\,s cutoff and over the instances that not every solver settles
within $5$\,s.

\clearpage
\section{Tableau Structure}\label{sec:tableau}
\noindent\resizebox{\textwidth}{!}{%
\begin{tikzpicture}[
  font=\sffamily,
  text=ink,
  row entry/.style={
    draw=rowblueborder,
    fill=rowblue,
    minimum width=2.55cm,
    minimum height=0.90cm,
    align=center,
    inner sep=3pt,
    font=\sffamily\small
  },
  column slot/.style={
    draw=colgreenborder,
    fill=colgreen,
    minimum width=1.35cm,
    minimum height=0.82cm,
    align=center,
    inner sep=3pt,
    font=\sffamily\small
  },
  linked entry/.style={draw=accentgold, line width=1.05pt},
  link marker/.style={
    circle,
    draw=accentgold,
    fill=white,
    text=accentgold,
    minimum size=4.5mm,
    inner sep=0pt,
    font=\sffamily\bfseries\scriptsize
  },
  array label/.style={font=\sffamily\bfseries\small, anchor=east},
  panel/.style={draw=black!28, fill=black!1, inner sep=8pt},
  operation arrow/.style={-{Stealth[length=2.2mm]}, draw=repairred, line width=1pt},
  operation box/.style={
    draw=black!38,
    fill=white,
    minimum height=0.78cm,
    align=center,
    inner sep=4pt,
    font=\sffamily\small
  }
]

\node[font=\sffamily\bfseries\large, anchor=west] (rows title) at (0,3.80)
  {Sparse basic rows};
\node[font=\sffamily\scriptsize, text=muted, anchor=west] at (0,3.30)
  {entries are triples $\langle x,c,i\rangle$, contiguous and sorted by $x$};

\node[array label] (r5 label) at (1.15,2.35) {\texttt{rows\_[x5]}};
\node[row entry, linked entry, anchor=west] (r5x0) at (1.40,2.35)
  {$x=x_0$, $c=2$\\[-1pt]\scriptsize $i=0$};
\node[row entry, right=0pt of r5x0] (r5x3)
  {$x=x_3$, $c=-1$\\[-1pt]\scriptsize $i=0$};

\node[array label] (r6 label) at (1.15,1.15) {\texttt{rows\_[x6]}};
\node[row entry, anchor=west] (r6x3) at (1.40,1.15)
  {$x=x_3$, $c=1$\\[-1pt]\scriptsize $i=1$};
\node[row entry, right=0pt of r6x3] (r6x4)
  {$x=x_4$, $c=7$\\[-1pt]\scriptsize $i=0$};

\node[array label] (r7 label) at (1.15,-0.05) {\texttt{rows\_[x7]}};
\node[row entry, linked entry, anchor=west] (r7x0) at (1.40,-0.05)
  {$x=x_0$, $c=1$\\[-1pt]\scriptsize $i=1$};
\node[row entry, right=0pt of r7x0] (r7x1)
  {$x=x_1$, $c=1$\\[-1pt]\scriptsize $i=0$};
\node[row entry, right=0pt of r7x1] (r7x4)
  {$x=x_4$, $c=-3$\\[-1pt]\scriptsize $i=1$};

\node[font=\sffamily\bfseries\large, anchor=west] (occ title) at (11.25,3.80)
  {Reverse occurrence index};
\node[font=\sffamily\scriptsize, text=muted, anchor=west] (occ subtitle) at (11.25,3.30)
  {$\occ[v]$: unsorted vector of rows containing $v$};

\node[array label] at (12.60,2.35) {\texttt{occ[x0]}};
\node[column slot, linked entry, anchor=west] (o0s0) at (12.85,2.35)
  {\scriptsize $i=0$\\$y=x_5$};
\node[column slot, linked entry, right=0pt of o0s0] (o0s1)
  {\scriptsize $i=1$\\$y=x_7$};

\node[array label] at (12.60,1.15) {\texttt{occ[x1]}};
\node[column slot, anchor=west] (o1s0) at (12.85,1.15)
  {\scriptsize $i=0$\\$y=x_7$};

\node[array label] at (12.60,-0.05) {\texttt{occ[x3]}};
\node[column slot, anchor=west] (o3s0) at (12.85,-0.05)
  {\scriptsize $i=0$\\$y=x_5$};
\node[column slot, right=0pt of o3s0] (o3s1)
  {\scriptsize $i=1$\\$y=x_6$};

\node[array label] at (12.60,-1.25) {\texttt{occ[x4]}};
\node[column slot, anchor=west] (o4s0) at (12.85,-1.25)
  {\scriptsize $i=0$\\$y=x_6$};
\node[column slot, right=0pt of o4s0] (o4s1)
  {\scriptsize $i=1$\\$y=x_7$};

\node[link marker, anchor=south] at ([yshift=-1.6mm]r5x0.north) {A};
\node[link marker, anchor=south] at ([yshift=-1.6mm]o0s0.north) {A};
\node[link marker, anchor=south] at ([yshift=-1.6mm]r7x0.north) {B};
\node[link marker, anchor=south] at ([yshift=-1.6mm]o0s1.north) {B};

\node[draw=accentgold, fill=accentgold!8, align=center, inner sep=6pt,
  text width=13.5cm, font=\sffamily\small] (invariant) at (7.9,-2.55)
  {if the row indexed by $y$ contains $\langle x,c,i\rangle$,
   then $\occ[x][i] = y$};

\node[font=\sffamily\bfseries\large, anchor=west] (repair title) at (0,-4.35)
  {Removing one occurrence by swap-pop};

\node[font=\sffamily\bfseries\small] at (5.10,-5.05) {1. Swap-pop the column};
\node[font=\sffamily\scriptsize] at (3.50,-5.60) {before};
\node[operation box, anchor=west, minimum width=1.30cm] (before0) at (2.20,-6.20)
  {$i=0$\\$\mathtt{x5}$};
\node[operation box, right=0pt of before0, minimum width=1.30cm] (before1)
  {$i=1$\\$\mathtt{x7}$};

\node[font=\sffamily\scriptsize] at (7.90,-5.60) {after};
\node[operation box, anchor=west, minimum width=1.30cm] (after0) at (7.25,-6.20)
  {$i=0$\\$\mathtt{x7}$};
\draw[operation arrow] (before1.east) -- (after0.west);
\node[font=\sffamily\scriptsize, text=repairred] at (5.85,-6.95)
  {unlink slot 0; move last};

\node[font=\sffamily\bfseries\small] at (16.05,-5.05) {2. Repair the moved row entry};
\node[font=\sffamily\scriptsize] at (13.15,-5.60) {before};
\node[operation box, text width=4.60cm] (repair before) at (13.15,-6.20)
  {\texttt{rows\_[x7]} entry for $x_0$\\$i=1$};
\node[font=\sffamily\scriptsize] at (19.65,-5.60) {after};
\node[operation box, text width=4.60cm] (repair after) at (19.65,-6.20)
  {\texttt{rows\_[x7]} entry for $x_0$\\$i=0$};
\draw[operation arrow] (repair before.east) -- (repair after.west);
\node[font=\sffamily\scriptsize, text=repairred] at (16.05,-6.95)
  {binary-search the sorted row: $O(\log |\mathtt{row}_{x7}|)$};

\node[font=\sffamily\scriptsize, text=muted, anchor=west] (repair note) at (0,-7.50)
  {The removed \texttt{rows\_[x5]} entry disappears; only the surviving moved
   column element needs a new back-index.};

\begin{scope}[on background layer]
  \node[panel, fit=(rows title)(r5 label)(r7x4)(r7 label)] {};
  \node[panel, fit=(occ title)(occ subtitle)(o0s0)(o0s1)(o4s0)(o4s1)] {};
  \node[panel, fill=repairred!3,
    fit=(repair title)(before0)(before1)(after0)(repair before)(repair after)(repair note)] {};
\end{scope}

\end{tikzpicture}%
}
 \end{document}